\documentclass[a4paper,11pt]{article}
\pdfoutput=1
\usepackage{jheppub}
\usepackage{mathtools}
\usepackage{physics}
\usepackage{orcidlink}
\graphicspath{{plots/}}

\allowdisplaybreaks

\newcommand{\gs}{g_\star}
\newcommand{\gss}{g_{\star s}}

\newcommand{\ogw}{\Omega_\text{GW}}
\newcommand{\calM}{\mathcal{M}}
\newcommand{\Msqr}[1]{\abs{\calM_{#1}}^2}

\title{Probing Bose-enhanced Inflaton Decay\\with Gravitational Waves}
\author[a]{Nicolás Bernal\,\orcidlink{0000-0003-1069-490X}}
\author[b,c]{Quan-feng Wu\,\orcidlink{0000-0002-5716-5266}}
\author[b]{Xun-Jie Xu\,\orcidlink{0000-0003-3181-1386}}
\author[d]{and Yong Xu\,\orcidlink{0000-0002-4582-8747}}
\affiliation[a]{New York University Abu Dhabi\\
PO Box 129188, Saadiyat Island, Abu Dhabi, United Arab Emirates}
\affiliation[b]{Institute of High Energy Physics, Chinese Academy of Sciences, Beijing 100049, China}
\affiliation[c]{Kaiping Neutrino Research Center, Kaiping 529386, China}
\affiliation[d]{McGill University Department of Physics \& Trottier Space Institute\\
3600 Rue University, Montréal, QC, H3A 2T8, Canada}
\emailAdd{nicolas.bernal@nyu.edu}
\emailAdd{wuquanfeng@ihep.ac.cn}
\emailAdd{xuxj@ihep.ac.cn}
\emailAdd{yong.xu6@mcgill.ca}

\abstract{We investigate cosmic reheating dynamics in the presence of a transient condensate formed by bosonic decay products of the inflaton. We show that the emergence of such a condensate and the corresponding Bose enhancement can dramatically increase the efficiency of inflaton decay, giving rise to qualitatively new reheating dynamics beyond the standard perturbative picture. As a consequence, graviton production from inflaton decay processes is significantly amplified by Bose enhancement effects, leading to a stochastic gravitational-wave background with a potentially observable amplitude, even in the low-frequency regime.}
\begin{document}
\begin{flushright}
\end{flushright}
\maketitle

%%%%%%%%%%%%%%%%%%%%%%%%%%%%%%%%%%%%%
\section{Introduction}
%%%%%%%%%%%%%%%%%%%%%%%%%%%%%%%%%%%%%
Gravitons are unique messengers from the very early Universe: once produced, they propagate essentially freely and retain detailed information about the cosmological background at their origin. Consequently, the spectrum of a stochastic gravitational-wave (GW) background provides a powerful and direct probe of primordial dynamics.

In inflationary cosmology, gravitons can be abundantly produced during the cosmic reheating epoch after inflation, when inflatons transfer their energy to the visible sector; for reviews of reheating, see, e.g., Refs.~\cite{Bassett:2005xm, Allahverdi:2010xz, Amin:2014eta, Lozanov:2019jxc, Lozanov:2020zmy, Barman:2025lvk}. Even in the minimal realization of reheating via perturbative inflaton decay, graviton production is unavoidable due to the universal coupling of gravity to energy--momentum of the matter content. Various processes contribute during this stage, including graviton bremsstrahlung from inflaton decay~\cite{Nakayama:2018ptw, Huang:2019lgd, Barman:2023ymn, Barman:2023rpg, Kanemura:2023pnv, Bernal:2023wus, Tokareva:2023mrt, Hu:2024awd, Choi:2024acs, Barman:2024htg, Inui:2024wgj, Jiang:2024akb, Datta:2025wfh, Das:2025cqs, Cheng:2025gyh}, inflaton pair annihilation~\cite{Ema:2015dka, Ema:2016hlw, Ema:2020ggo, Choi:2024ilx, Mudrunka:2026kgm}, scattering between the inflaton and its decay products~\cite{Xu:2024fjl, Bernal:2025lxp, Xu:2025wjq}, and purely scattering processes among the daughter particles~\cite{Bernal:2024jim, Xu:2025wjq}.

However, a generic challenge arises from the Planck suppression of gravitational interactions. Achieving a sizable GW amplitude typically requires a high characteristic energy scale, pushing the peak of the GW spectrum into the ultra-high-frequency regime and favoring large inflaton masses and/or high reheating temperatures. This substantially limits the detectability of GWs generated during reheating, given the current lack of sufficiently sensitive high-frequency detectors~\cite{Aggarwal:2020olq, Aggarwal:2025noe}. 

Recently, we demonstrated that this limitation can be overcome if the inflaton scatters with a non-thermalized bosonic decay product, dubbed the \emph{reheaton}~\cite{Bernal:2025lxp}. During the pre-thermalization stage, the reheaton occupation number can become parametrically large, enhancing graviton production and allowing for an observable GW signal even for comparatively small inflaton masses. This enhancement simultaneously shifts the GW spectrum toward lower frequencies, improving the detection prospects.

In this work, we uncover a qualitatively new and even more powerful realization of this idea with transient \emph{reheaton condensation}. We show that when the reheaton forms a condensate, quantum-statistical enhancement dramatically alters the inflaton decay dynamics and opens entirely new, highly efficient channels for graviton production. This mechanism leads to GW signals with amplitudes and spectral features that are markedly distinct from those of previously studied reheating scenarios, providing a novel observational window into the microscopic physics of reheating.

The remainder of this work is organized as follows. In Section~\ref{sec:formulation}, we present the formalism of inflaton decay, with particular emphasis on Bose enhancement effects. The resulting GW spectra are discussed in Section~\ref{sec:GWs}. Our main conclusions are summarized in Section~\ref{sec:conclusion}. Technical details of the phase-space integrations and computations for collision terms are relegated to the Appendices.

%%%%%%%%%%%%%%%%%%%%%%%%%%%%%%%%%%%%%%%%%%%%%%%%%%%%%%%%%%
\section{Reheating through inflaton decay \label{sec:formulation}}
%%%%%%%%%%%%%%%%%%%%%%%%%%%%%%%%%%%%%%%%%%%%%%%%%%%%%%%%%%
In this section, the cosmic reheating era is studied in the context of a perturbative decay of the inflaton field.\footnote{This simplified setup is adopted to illustrate the underlying physical mechanism. For non-perturbative reheating scenarios, including preheating, we refer the reader to the reviews in Refs.~\cite{Allahverdi:2010xz, Amin:2014eta, Lozanov:2019jxc, Barman:2025lvk}.} Inflatons decay into pairs of light mediator particles (the reheatons), which are assumed to have no sizable self-interactions, such that they only redshift without reaching kinetic or chemical equilibrium. We first review the standard scenario where the decay proceeds without Bose enhancement, followed by the case of a Bose-enhanced inflaton decay. Eventually, at late times, all reheatons transfer their energy to SM particles, while still being relativistic.

%%%%%%%%%%%%%%%%%%%%%%%%%%%%%%%%%%%%%%%%%%%%%%%%%%%%%%%%%%
\subsection{Inflaton decay without Bose enhancement\label{subsec:decay-no-enhance}}
\begin{figure}
    \centering
    \includegraphics[width=0.85\textwidth]{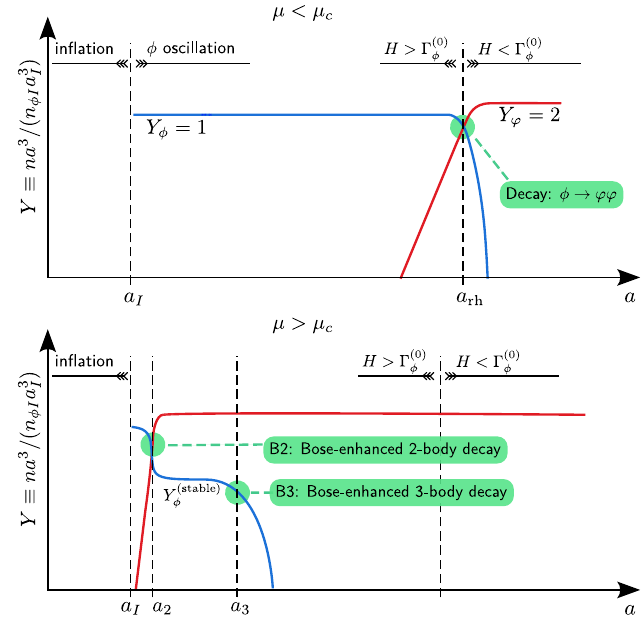}
    \caption{Schematic illustration of the evolution of the normalized comoving number densities $Y$ of inflatons ($\phi$, blue lines) and reheatons ($\varphi$, red lines) with (bottom panel, $\mu \gg \mu_c$) and without (top panel, $\mu \ll \mu_c$) Bose enhancement, as a function of the cosmic scale factor $a$.
    The vertical dashed lines correspond to different characteristic epochs in the evolution of the background. $a=a_I$ is the beginning of the reheating era. For $\mu < \mu_c$ (top), the inflaton effectively decays into a pair of reheatons at $a=a_\text{rh}$. For $\mu > \mu_c$ (bottom), the inflaton first rapidly decays through a Bose-enhanced 2-body decay $\phi \to \varphi \varphi$ at $a= a_2$, while at $a= a_3$ the Bose-enhanced 3-body decay $\phi \to \varphi \varphi h$ becomes relevant.
    }
    \label{fig:schematic}
\end{figure}

%%%%%%%%%%%%%%%%%%%%%%%%%%%%%%%%%%%%%%%%%%%%%%%%%%%%%%%%%%
Let us start with one of the simplest perturbative reheating scenarios: inflaton decay. After inflation, the inflaton field $\phi$ oscillates at the bottom of its potential, which we assume to be quadratic. This leads to a matter-dominated epoch during which the Universe is filled with cold $\phi$ particles. If $\phi$ is assumed to have a small perturbative decay width, $\phi\to\varphi\varphi$, where $\varphi$ is a light relativistic boson called the reheaton throughout this work, all $\phi$ particles will gradually decay, releasing their energy to $\varphi$, resulting in a radiation-dominated era. This is illustrated in the upper panel of Fig.~\ref{fig:schematic}, where the blue and red lines represent the normalized comoving number densities $Y$ of $\phi$ and $\varphi$, respectively. 
%%%%%%%%%%%%%%%%%%%%%%%%%%%%%%%%%%%%%%%%%%%%%%%%%%%%%%%%%%
%%%%%%%%%%%%%%%%%%%%%%%%%%%%%%%%%%%%%%%%%%%%%%%%%%%%%%%%%%

Quantitatively, it is possible to calculate this reheating process, starting from the relevant Lagrangian density
\begin{equation}
    {\cal L} \supset - \frac12\, m_\phi^2\, \phi^2 - \frac{\mu}{2}\, \phi\, \varphi^2\thinspace,\label{eq:L}
\end{equation}
where $m_\phi$ denotes the mass of the inflaton and $\mu$ is a dimension-one coupling.  Here, the bare mass term for the reheaton is ignored, as it is assumed to be always relativistic. In addition, as we focus on a non-thermal reheaton, possible trilinear and quartic self-interactions are assumed to be subdominant. For later convenience, we also define the dimensionless coupling
\begin{equation}
    y \equiv \frac{\mu}{m_\phi}\,.\label{eq:}
\end{equation}
From the Feynman diagrammatic perspective, the theory is perturbative as long as $y^2 \ll 4 \pi$. The tree-level decay rate of $\phi\to\varphi\varphi$ reads 
\begin{equation}
    \Gamma_\phi^{(0)} \simeq \frac{\mu^2}{32 \pi\, m_\phi} = \frac{y^2}{32\pi}\, m_\phi\thinspace.\label{eq:-1}
\end{equation}
Note that $\Gamma_\phi^{(0)}$ corresponds to the {\it in-vacuum} decay rate. In the early Universe with a dense background of particles, the actual decay rate, denoted by $\Gamma_\phi$, may be significantly different from $\Gamma_\phi^{(0)}$ in Eq.~\eqref{eq:-1}. Nevertheless, if the decay occurs sufficiently slow (that is, for $\mu$ sufficiently small), the difference can be safely ignored, and Eq.~\eqref{eq:-1} can be used. This statement will be quantified in the next section. 

The number densities of $\phi$ and $\varphi$, denoted by $n_\phi$ and $n_\varphi$, are governed by the following set of Boltzmann equations
\begin{align}
    \frac{dn_\phi}{dt} + 3\, H\, n_\phi & = -\Gamma_\phi\, n_\phi\thinspace,\label{eq:-2}\\
    \frac{dn_\varphi}{dt} + 3\, H\, n_\varphi & = +2\, \Gamma_\phi\, n_\phi\thinspace,\label{eq:-3}
\end{align}
where $H \equiv \dot{a}/a$ is the Hubble expansion rate with $a$ the cosmic scale factor. It is determined by
\begin{equation}
    H^2 = \frac{\rho_\phi + \rho_\varphi}{3\, M_P^2}\thinspace,\label{eq:-4}
\end{equation}
where $\rho_{\phi,\varphi}$ denotes the energy density of $\phi$ and $\varphi$, and $M_P\equiv 1/\sqrt{8 \pi\, G_N} \simeq 2.435 \times 10^{18}$~GeV, with $G_N$ being Newton's gravitational constant, is the reduced Planck mass.  In Eq.~\eqref{eq:-3}, we assume that $\varphi$ is produced only through $\phi \to \varphi \varphi$, while other processes that could potentially change the number of $\varphi$ particles (e.g. $\varphi \to \text{SM}$, $\phi \to 4 \varphi$, $2 \varphi \to 4 \varphi$) are highly suppressed.

Taking $\Gamma_\phi = \Gamma_\phi^{(0)}$, Eqs.~\eqref{eq:-2} and~\eqref{eq:-3} can be analytically solved from the beginning of reheating at $a = a_I$. Taking $n_\phi(a_I) \equiv n_{\phi I} = 3 H_I^2\, M_P^2/m_\phi$ and $n_\varphi(a_I) = 0$  as initial conditions, one gets~\cite{Bernal:2025lxp}
\begin{align}
    n_\phi(a) & = n_{\phi I} \left(\frac{a_I}{a}\right)^3 \exp\left[-\Gamma_\phi^{(0)}\, t\right],\\
    n_\varphi(a) &= 2 \left[n_{\phi I} \left(\frac{a_I}{a}\right)^3 - n_\phi(a)\right],
\end{align}
with
\begin{equation}
    t\simeq\frac{2}{3\, H_I} \left[\left(\frac{a}{a_I}\right)^{3/2} - 1\right].\label{eq:-7}
\end{equation}
The evolution of the normalized comoving number densities
\begin{equation}
    Y_i(a) \equiv \frac{n_i}{n_{\phi I}} \left(\frac{a}{a_I}\right)^3
\end{equation}
with $i = \phi$ and $\varphi$ is schematically illustrated in the upper panel of Fig.~\ref{fig:schematic}.

Before ending this section, we note that  the trilinear inflaton–reheaton coupling $\mu$ in Eq.~\eqref{eq:L} can induce an effective mass term for the latter, proportional to the inflaton field value~$\phi$, thus modifying the decay kinematics~\cite{Ichikawa:2008ne, Drewes:2019rxn, Garcia:2020wiy}. To properly account for this effect, one should average over the inflaton oscillations, which leads to an effective coupling $\mu_{\text{eff}}$ for the interaction in Eq.~\eqref{eq:-1}. For the quadratic inflaton potential considered here, this effect has been shown to be moderate, yielding $\mu_{\text{eff}} \simeq \mu$~\cite{Ichikawa:2008ne, Garcia:2020wiy}. Moreover,  this coupling could also lead to a tachyonic mass for the daughter field when the inflaton crosses zero during its oscillations, potentially triggering nonperturbative particle production via tachyonic resonance~\cite{Dufaux:2006ee}. However, even a small self-interaction of the daughter field, such as a $\lambda\, \varphi^4$ term, generates a positive effective mass-squared contribution of order $\lambda\, \langle \varphi^2 \rangle$, which counteracts the tachyonic instability. Here, $\langle \varphi^2 \rangle$ denotes the field variance, i.e., the expectation value of the squared reheaton field, which grows as particles are produced and provides a dynamical mass that stabilizes the system. As a result, nonperturbative energy transfer becomes subdominant due to backreaction, and the inflaton energy is primarily transferred via perturbative decay. In this work, we restrict our analysis to the simplest perturbative scenario and illustrate the effectiveness of Bose enhancement in draining the inflaton energy even in the perturbative regime.

%%%%%%%%%%%%%%%%%%%%%%%%%%%%%%%%%%%%%%%%%%%%%%%%%%%%%%%%%%
\subsection{Inflaton decay with Bose enhancement}
%%%%%%%%%%%%%%%%%%%%%%%%%%%%%%%%%%%%%%%%%%%%%%%%%%%%%%%%%%
The treatment in Section~\ref{subsec:decay-no-enhance} is valid as long as the decay of $\phi$ is not affected by the cosmological background. However, if $\phi$ decays fast, the number density of the decay product, $\varphi$, can increase rapidly to a level at which its influence on the decay can no longer be ignored.

Qualitatively, one can anticipate that the decay is enhanced by the presence of the $\varphi$ background due to its {\it bosonic} nature. As $\mu$ increases, the Bose enhancement effect is enhanced due to the rapidly-created $\varphi$ background. Quantitatively, this can be calculated from the following unintegrated Boltzmann equation
\begin{equation}
    \left(\frac{\partial}{\partial t} - H\, p\, \frac{\partial}{\partial p}\right) f(t,p) = {\cal C}[f]\thinspace,\label{eq:f-boltz}
\end{equation}
where $f(t,p)$ is the phase-space distribution function of the species under consideration and ${\cal C}[f]$ is the collision term. The most general form of ${\cal C}[f]$ is well known and can be found in cosmology textbooks (e.g., see Ref.~\cite{Kolb:1990eaun}). After performing the phase space integration in ${\cal C}[f]$ (see Appendix~\ref{sec:integral} for the detailed derivation), we obtain
\begin{align}
    {\cal C}[f_\phi] &= -\Gamma_\phi^{(0)}\, f_\phi\, {\cal B}_e\thinspace,\label{eq:-9}\\
    {\cal C}[f_\varphi] &= +\frac{16\pi^2\, \Gamma_\phi^{(0)}\, n_\phi}{m_\phi^2}\, \delta\left(p_\varphi - \frac{m_\phi}{2}\right)\, {\cal B}_e\thinspace,\label{eq:-10}
\end{align}
where the Bose-enhancement factor ${\cal B}_e$ is given by
\begin{equation}
    {\cal B}_e \equiv 1 + f_\varphi \left(t,\frac{m_\phi}{2}\right).\label{eq:-11}
\end{equation}
If we further integrate out the phase space of $\phi$ or $\varphi$, Eqs.~\eqref{eq:-9} and~\eqref{eq:-10} lead to
\begin{align}
    \int {\cal C}[f_\phi]\, \frac{4\pi\, p_\phi^2\, dp_\phi}{(2\pi)^3} &= -n_\phi\, \Gamma_\phi^{(0)}\, {\cal B}_e\thinspace,\label{eq:-12}\\
    \int {\cal C}[f_\varphi]\, \frac{4\pi\, p_\varphi^2\, dp_\varphi}{(2\pi)^3} &= + 2\, n_\phi\, \Gamma_\phi^{(0)}\, {\cal B}_e\thinspace.\label{eq:-13}
\end{align}
This should be compared with the right-hand sides of Eqs.~\eqref{eq:-2} and~\eqref{eq:-3}, implying 
\begin{equation}
    \Gamma_\phi = \Gamma_\phi^{(0)}\, {\cal B}_e\thinspace.\label{eq:-14}
\end{equation}
Obviously, in the absence of Bose enhancement, ${\cal B}_e\to1$, $\Gamma_\phi$ becomes identical to $\Gamma_\phi^{(0)}$, while ${\cal C}[f_\phi]$ and ${\cal C}[f_\varphi]$ recover their respective forms in the case of Maxwell-Boltzmann statistics. 

A highly non-trivial feature of the above formalism including Bose enhancement is the following series of consequences: 
\begin{equation*}
    \boxed{\text{increasing }f_\varphi\to\text{increasing }{\cal B}_e\to\text{enhanced decay rate}\to\text{increasing }f_\varphi\thinspace.}
\end{equation*}
The first and second steps can be immediately seen from Eqs.~\eqref{eq:-11} and~\eqref{eq:-14}. The last step is valid until the remaining $\phi$ particles  are insufficient  to support Bose-enhanced decay. Due to this closed positive-feedback loop, the system exhibits strongly nonlinear behavior. Under certain conditions, the feedback loop can lead to instability, causing $\phi$ to decay at an extremely rapid rate. 

In the following, we study this nonlinear behavior analytically. Substituting Eq.~\eqref{eq:-10} into Eq.~\eqref{eq:f-boltz} and assuming that the evolution of $n_\phi$ is known, one can solve the Boltzmann equation analytically (see Appendix~\ref{sec:solve-f-R}) and obtain
\begin{equation}
    f_\varphi = \exp\left[\frac{\pi\, \mu^2}{m_\phi^4}\, \frac{n_\phi(a\, x_\varphi)}{H(a\, x_\varphi)}\, \Theta\left[\frac{a_I}{a} \leq x_\varphi \leq 1\right]\right] - 1\thinspace,\label{eq:f-reheaton}
\end{equation}
where 
\begin{equation}
    x_\varphi \equiv \frac{2\, p_\varphi}{m_\phi}\thinspace.\label{eq:xp}
\end{equation}
Substituting Eq.~\eqref{eq:f-reheaton} into Eq.~\eqref{eq:-11}, we obtain
\begin{equation}
    {\cal B}_e = \exp\left[\frac{\pi\, \mu^2}{m_\phi^4}\, \frac{n_\phi(a)}{H(a)}\right].\label{eq:Be-exp}
\end{equation}

It is convenient to rewrite Eq.~\eqref{eq:-2} as
\begin{equation}
    \frac{1}{a^3}\, \frac{d(n_\phi\,a^3)}{da} = -\frac{\Gamma_\phi^{(0)}}{a\, H}\, {\cal B}_e\, n_\phi\,,\label{eq:-18}
\end{equation}
which is equivalent to
\begin{equation} \label{eq:dYda}
    \frac{dY_\phi}{da} = -\beta\, Y_\phi\,,
\end{equation}
where
\begin{equation}\label{eq:beta}
    \beta \equiv \frac{\Gamma_\phi^{(0)}}{a\, H}\, \exp\left[\frac{\pi\, \mu^2\, n_{\phi I}}{m_\phi^4\, H} \left(\frac{a_I}{a}\right)^3 Y_\phi\right],
\end{equation}
with $\beta$ being the depletion rate of $Y_\phi$. The exponential dependence on the trilinear coupling $\mu$ in $\beta$ has an important implication: if the exponent exceeds a critical value, inflatons decay almost instantaneously, exponentially fast. This exponential behavior has also been investigated previously in Refs.~\cite{Moroi:2020has, Moroi:2020bkq}.

\begin{figure}
    \centering
    \includegraphics[width=0.6\textwidth]{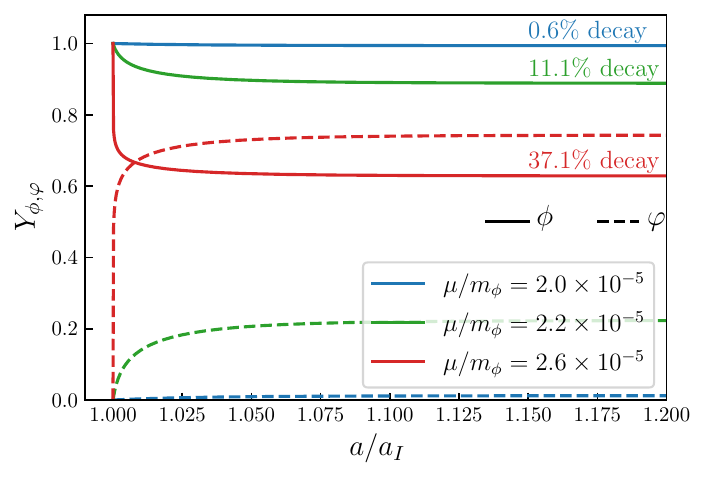}
    \caption{\label{fig:critical-mu} Numerical examples showing how the evolution of the normalized comoving number densities for $\phi$ and $\varphi$ are strongly affected by small variations of $\mu$ around the critical value $\mu_c$. It is assumed $m_\phi = 10^{13}$~GeV and $H_I/m_\phi = 0.1$.}
\end{figure}

To illustrate this exponential sensitivity, we inspect several numerical examples. For $m_\phi = 10^{13}$~GeV, $H_I/m_\phi = 0.1$ and $\mu/m_\phi = \{2,\,3,4\} \times 10^{-5}$, the corresponding values of $\Gamma_\phi^{(0)}$ and $\beta$ are $\Gamma_\phi^{(0)} \simeq \{4, 9, 16\} \times 10^{-12}\, m_\phi \ll H_I$ and $\beta \simeq \{0.2, 6 \times 10^{11}, 10^{29}\}$. The last two cases indicate that $Y_\phi$ should decrease almost instantly. To further inspect how the evolution of $Y_\phi$ is affected by the Bose enhancement, we present numerical solutions of Eq.~\eqref{eq:dYda} with $\mu/m_\phi = \{2.0, 2.2, 2.6\} \times 10^{-5}$ in Fig.~\ref{fig:critical-mu}. The case $\mu/m_\phi = 2 \times 10^{-5}$ leads to a very insignificant decay (only $0.6\%$) of inflatons. When $\mu/m_\phi$ increases to $2.2 \times 10^{-5}$, a significant portion ($11.1\%$) of the inflatons decay rapidly due to the Bose enhancement. The decrease in $Y_\phi$ by $11.1\%$ simply reflects the increase in $\sim 10\%$ of $\mu$. However, increasing further $\mu/m_\phi$ to $2.6 \times 10^{-5}$ translates into a rapid decay of more than a third of the inflatons, producing a large number of reheatons with $n_\varphi > n_\phi$. We emphasize that this behavior corresponds to the first depletion of inflatons shown in the lower panel of Fig.~\ref{fig:schematic}, and that eventually, at later times, all remaining inflatons will decay.
%%%%%%%%%%%%%%%%%%%%%%%%%%%%%%%%%%%%%%%%%%%%%%%%%%%%%%%%%%%%%%
%%%%%%%%%%%%%%%%%%%%%%%%%%%%%%%%%%%%%%%%%%%%%%%%%%%%%%%%%%%%%%

Since the condition required for rapid depletion of $Y_\phi$ is $\beta\gg1$, one can use it to derive the critical value of $\mu$ for this to occur, together with the final stable value of $Y_\phi$ after depletion. As $Y_\phi$ decreases, the exponent in Eq.~\eqref{eq:dYda} can decrease to a level that cannot maintain $\beta \gg 1$, effectively inhibiting $Y_\phi$ from further decreasing. Solving $\beta=1$ with respect to $Y_\phi$ and assuming that the variation of $H$ is small during this rapid process (i.e., $H\simeq H_I$), we obtain
\begin{equation}
    Y_\phi^{({\rm stable})} \simeq \frac{H_I\, m_\phi^4}{\pi\, \mu^2\, n_{\phi I}}\, \ln\left(\frac{32\pi\, H_I\, m_\phi}{\mu^2}\right), \label{eq:Y-phi-stable}
\end{equation}
which is the stable value of $Y_\phi$ after rapid depletion. In addition, $\beta=1$ can be translated to a critical value of $\mu$, denoted by $\mu_c$, given by
\begin{equation} \label{eq:muc}
    \mu_c^2 = \frac{m_\phi^5}{3 \pi\, M_P^2\, H_I}\, \mathcal{W}_0\left[\frac{96\pi^2\, M_P^2\, H_I^2}{m_\phi^4}\right],
\end{equation}
with $\mathcal{W}_0$ the principal branch of the Lambert function. For the benchmark values of $m_\phi$ and $H_I$ used in Fig.~\ref{fig:critical-mu}, we have $\mu_c/m_\phi \simeq 2.07\times10^{-5}$, implying that the orange and green lines have $\mu>\mu_c$. According to Eq.~\eqref{eq:Y-phi-stable}, the corresponding values of $Y_\phi^{({\rm stable})}$ are $0.88$ and $0.62$, in good agreement with the numerical results in Fig.~\ref{fig:critical-mu}. 

\begin{figure}
    \def\sepf{0.481}
    \centering
    \includegraphics[scale=\sepf]{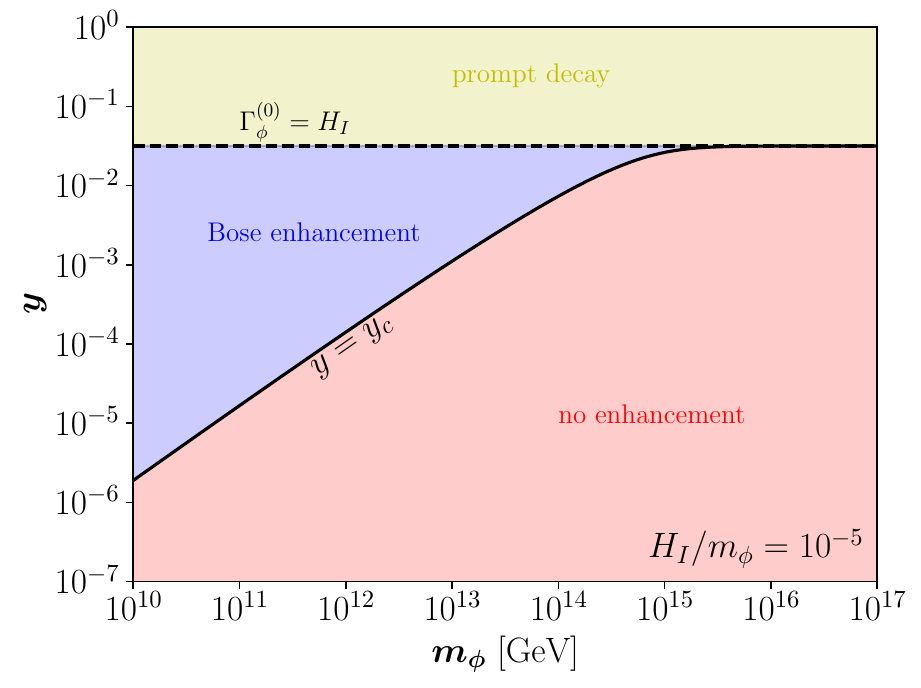}
    \includegraphics[scale=\sepf]{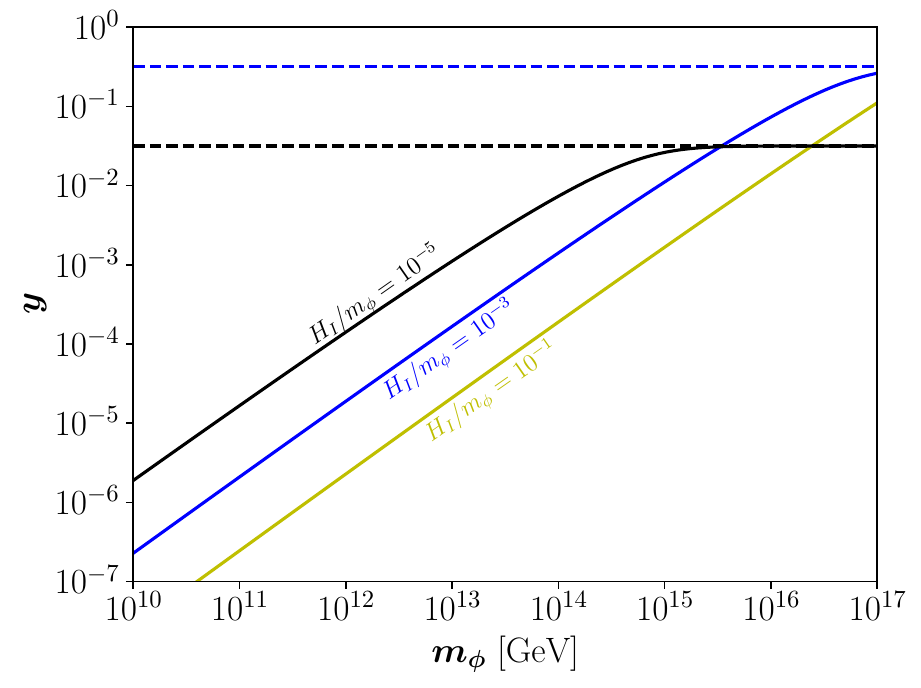}
    \caption{Critical valued of $y_c \equiv \mu_c/m_\phi$ (solid lines) and $\Gamma_\phi^{(0)} = H_I$ (dashed lines), for different values of $H_I/m_\phi = 10^{-5}$ (black), $H_I/m_\phi = 10^{-3}$ (blue), or $H_I/m_\phi = 10^{-1}$ (yellow). For $y < y_c$ (red area) the inflaton decays without a sizable Bose enhancement, while for $\Gamma_\phi^{(0)} > H_I$ (yellow area) its decay is prompt even without Bose enhancement. In between, inflatons have a Bose-enhanced decay (blue area).}
    \label{fig:y-bound}
\end{figure}

In the left panel of Fig.~\ref{fig:y-bound}, the black solid line shows the critical value of $y_c \equiv \mu_c/m_\phi$ as a function of the inflaton mass $m_\phi$, assuming $H_I/m_\phi = 10^{-5}$. Above this line, Bose enhancement leads to exponentially fast decay of the inflaton (cf.~the lower panel of Fig.~\ref{fig:schematic}), whereas below, the inflaton decays are non-enhanced (cf.~the upper panel of Fig.~\ref{fig:schematic}). In the small-mass regime, we find $\mu_c^2 \propto m_\phi^5 /(M_P^2 H_I)$, up to a mild logarithmic correction arising from $\mathcal{W}_0$ in Eq.~\eqref{eq:muc}, implying $y_c \propto m_\phi^{3/2}/(M_P H_I^{1/2})$. As $m_\phi$ increases, $\mathcal{W}_0\!\left(96\pi^2 M_P^2 H_I^2/m_\phi^4\right) \to 96\pi^2 M_P^2 H_I^2/m_\phi^4$, and consequently $\mu_c^2 \to 32\pi H_I m_\phi$. As a result, $y_c \to (32 \pi H_I/m_\phi)^{1/2} = (H_I/\Gamma_\phi^{(0)})^{1/2}$, as indicated by the horizontal dashed line labeled $\Gamma^{(0)} = H_I$. Above this line, inflaton decay promptly, in a time shorter than a Hubble time, even without Bose enhancement. The right panel of Fig.~\ref{fig:y-bound} is similar to the left panel, 
but overlays the cases $H_I/m_\phi = 10^{-1}$ and $H_I/m_\phi = 10^{-3}$.
%%%%%%%%%%%%%%%%%%%%%%%%%%%%%%%%%%%%%%%%%%%%%%%%%%%%%
%%%%%%%%%%%%%%%%%%%%%%%%%%%%%%%%%%%%%%%%%%%%%%%%%%%%%
 
In this work, we focus on the scenario with $\mu>\mu_c$, i.e., the lower panel of Fig.~\ref{fig:schematic}. It is interesting to note that after the Bose-enhanced two-body decay $\phi\to\varphi\varphi$ (B2) effectively stops and $Y_\phi$ reaches $Y_\phi^{({\rm stable})}$ given by Eq.~\eqref{eq:Y-phi-stable}, another Bose-enhanced decay process, $\phi\to\varphi\varphi h$ with $h$ the graviton, may take over and further deplete inflatons. This is represented by ``B3: Bose-enhanced 3-body decay'' in the lower panel of Fig.~\ref{fig:schematic} and will be discussed in detail in the next section. 

Before closing this section, two comments are in order. First, our analysis assumes a quadratic potential, for which the inflaton oscillates with a fixed frequency $m_\phi$ and behaves as non-relativistic matter with $n_\phi=\rho_\phi/m_\phi$. For a quartic potential $V(\phi)\propto \phi^4$, the oscillation is anharmonic and the notion of a constant inflaton mass is no longer applicable. Instead, the relevant energy scale entering the decay kinematics is set by the instantaneous oscillation scale, which decreases with time due to Hubble expansion. Consequently, the Bose-enhancement factor in Eq.~\eqref{eq:Be-exp} acquires a different time dependence, modifying the onset condition for exponential growth and shifting the numerical value of the critical coupling $\mu_c$. This also changes the time at which graviton production is most efficient, and therefore shifts the overall normalization and characteristic frequency of the GW spectrum. However, the feedback mechanism encoded in Eqs.~\eqref{eq:Be-exp} to~\eqref{eq:beta} remains intact, so the existence of a Bose-enhanced regime and the associated production channels are unchanged.

Second, we have neglected self-interactions of $\varphi$. A quartic coupling $\lambda\, \varphi^4$ induces an effective mass that grows with the occupation number. As long as this induced mass remains smaller than the typical momentum of the produced particles, the reheatons remain relativistic and the decay kinematics are unchanged. In this regime, self-scattering is also subdominant to Hubble expansion, so the distribution remains non-thermal and the Bose-enhancement mechanism proceeds as described above.  For larger values of the self-coupling, self-interactions can broaden the momentum distribution, leading to a moderate reduction of the occupation number at the characteristic production scale. This results in a quantitative suppression of the Bose-enhancement factor and hence of the GW amplitude. However, the underlying instability and the associated graviton production channels remain intact, so the qualitative features of the signal are preserved.

%%%%%%%%%%%%%%%%%%%%%%%%%%%%%%%%%%%%%%%%%%%%%%%%%%%%%%%%%%%%%%%%%%%%%
\section{Production of gravitational waves \label{sec:GWs}}
%%%%%%%%%%%%%%%%%%%%%%%%%%%%%%%%%%%%%%%%%%%%%%%%%%%%%%%%%%%%%%%%%%%%%
The Bose-enhanced decay of the inflaton not only alters the reheating dynamics, as investigated in the previous section, but also leads to novel GW signatures. In this section, we compute the production of gravitons ($h$) from the following processes\footnote{For the Feynman diagrams of these processes and their corresponding amplitudes and production rates, we refer to Ref.~\cite{Bernal:2025lxp}.}
\begin{align*}
    \varphi\, \varphi & \to h\, h\thinspace,\\
    \phi & \to \varphi\, \varphi\,  h\thinspace,\\
    \varphi\, \phi & \to \varphi\,  h\thinspace,\\
    \phi\, \phi & \to h\, h\thinspace,
\end{align*}
in which inflatons and reheatons can self-annihilate, coannihilate, or decay. We consider all possible tree-level diagrams with at most four external particles that include at least one graviton. In addition, throughout this section, we assume $\mu>\mu_c$. Furthermore, we emphasize that the reheaton eventually decays into SM particles, with an interaction rate 
% low-enough so that the decay is late 
low enough to ensure that the decay occurs late 
and that the reheatons remain non-thermalized until the end of reheating.

As preliminary general comments, we note that the first process $\varphi\varphi\to hh$ tends to be independent of $\mu$ as long as $\mu \gg \mu_c$. The other three processes depend on the population of the residual particles $\phi$, which is roughly proportional to $\mu^{-2}$, as suggested by Eq.~\eqref{eq:Y-phi-stable}. For $\phi\to\varphi\varphi h$ and $\varphi\phi\to\varphi h$, although both have $\varphi$ in the final states, only the former can be Bose enhanced. The latter produces $\varphi$ with $p_\varphi$ always above $m_\phi/2$ while the momentum distribution of the background particles $\varphi$ peaks at $p_\varphi\simeq\frac{m_\phi}{2}\frac{a_I}{a}$. 

Since all of these processes are suppressed by the Planck mass scale, they are far from reaching equilibrium. Hence, their impact on the major component of the background (i.e., reheatons) is negligible. However, as we will show later, the residual population of inflatons is further reduced because of its Bose-enhanced three-body decay. 

%%%%%%%%%%%%%%%%%%%%%%%%%%%%%%%%%%%%%%%%%%%%%%%%%
\subsection{General formalism}
%%%%%%%%%%%%%%%%%%%%%%%%%%%%%%%%%%%%%%%%%%%%%%%%%
The production of gravitons from the processes mentioned above can be computed by solving the Boltzmann equation \eqref{eq:f-boltz} for the graviton phase-space distribution $f_h.$ Since $f_h \ll 1$ is always satisfied in this work, one can safely ignore all terms proportional to $f_h$ and define the production rate $\Gamma_h$
\begin{equation}
    \Gamma_h(a,p_h)\equiv\lim_{f_h\to0}{\cal C}[f_h]\thinspace.\label{eq:-8}
\end{equation}
This allows us to rewrite Eq.~\eqref{eq:f-boltz} into the integral form as~\cite{Bernal:2025lxp}
\begin{equation}
    f_h(a,p_h)=\int_{a_I}^{a}\frac{da'}{a'\,H(a')}\,\Gamma_h\left(a',p_h\,\frac{a}{a'}\right). \label{eq:f-from-int}
\end{equation}

In the present Universe, the produced gravitons are highly red-shifted, forming a GW background with the following energy density and differential energy density
\begin{align}
    \rho_\text{GW} & =g_h \int\frac{4\pi\, \omega^3\, d\omega}{(2\pi)^3}\, f_h(a_0,\omega)\,, \label{eq:rhoGW-int}\\
    \frac{d\rho_\text{GW}}{d\omega} & = g_h\, \frac{4\pi\, \omega^3}{(2\pi)^3}\, f_h(a_0,\omega)\thinspace,\label{eq:drhoGW}
\end{align}
where $\omega$ is the graviton energy, $g_h=2$ denotes the two degrees of freedom for massless gravitons, and $a_0$ is the scale factor at present. It is customary to express the differential energy density in terms of $\ogw$ which is defined as
\begin{equation}
    \ogw(f)\equiv\frac{1}{\rho_c}\,\frac{d\rho_\text{GW}}{d\ln\omega}=16\pi^2\,\frac{f^4}{\rho_c}\,f_h(a_0,2\pi\,f)\,,\label{eq:-19}
\end{equation}
where  $f \equiv \frac{\omega}{2\pi}$ is the GW frequency and $\rho_c \simeq 1.05 \times 10^{-5}~h^2$~GeV/cm$^3$ is the critical energy density at present~\cite{ParticleDataGroup:2024cfk}. At high frequencies, GWs are mainly constrained by the effective number of neutrino species ($N_{\rm eff}$) from the CMB and BBN observations. The contribution of GWs to $N_{\rm eff}$, denoted by $\Delta N_{\rm eff}$, is proportional to $\rho_\text{GW}$ and is given by
\begin{equation}
    \Delta N_{\rm eff} = \frac{3}{\Omega_\nu}\, \frac{\rho_\text{GW}}{\rho_c}\thinspace,\label{eq:-20}
\end{equation}
with $\Omega_\nu \equiv \rho_\nu/\rho_c \simeq 1.7 \times 10^{-5}/h^2$, where $\rho_\nu$ is the energy density of the cosmic neutrino background assuming massless neutrinos. Therefore, a bound on $\Delta N_{\rm eff}$ can be recast as a bound on $\ogw$ by~\cite{Caprini:2018mtu}
\begin{equation}
    \ogw \equiv \frac{\rho_\text{GW}}{\rho_c} = \frac{\Omega_\nu}{3}\, \Delta N_{\rm eff} \simeq 5.7 \times 10^{-6}\, h^{-2}\, \Delta N_{\rm eff} \thinspace\cdot\label{eq:-21}
\end{equation}
A subtle issue regarding the consistency between $\ogw \equiv \rho_\text{GW}/\rho_c$ used in Eq.~\eqref{eq:-21} and $\ogw(f) \equiv \frac{1}{\rho_c}\, \frac{d\rho_\text{GW}}{d\ln\omega}$ used in Eq.~\eqref{eq:-19} is to be clarified here. The latter is essentially a differential distribution of GW energies with respect to $\ln\omega$ or $\ln f$ while the former concerns the total energy density. Strictly speaking, the two $\ogw$ are different and should be represented by different notation (e.g, $d\ogw/d\ln f$ versus $\ogw$). However, following the convention in the literature, we use the same notation for both and keep in mind that the two can be distinguished by whether they are $f$-dependent. In addition, for a GW spectrum with an ${\cal O}(1)$ span of $\ln f$, we roughly have  $\int\ogw(f)\thinspace d\ln f \sim {\cal O}(1) \times \ogw$, which implies that $\ogw(f)$ and $\ogw$ can be interchangeably used up to an ${\cal O}(1)$ factor. 

%%%%%%%%%%%%%%%%%%%%%%%%%%%%%%%%%%%%%%%%%%%%%%%
\begin{figure}
    \centering
    \def\sepf{0.488}
    \centering
    \includegraphics[scale=\sepf]{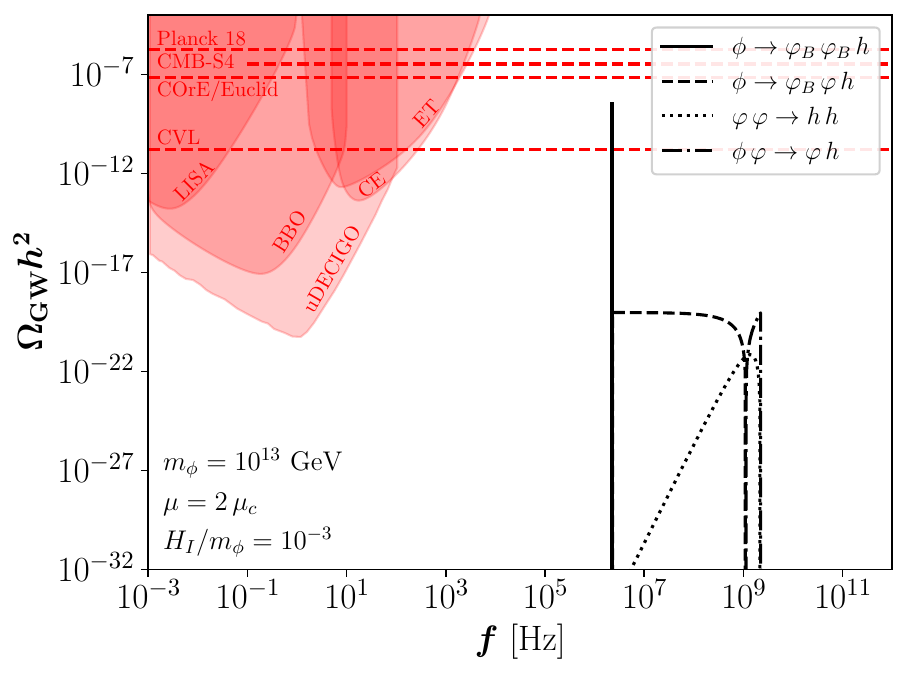}\\
    \includegraphics[scale=\sepf]{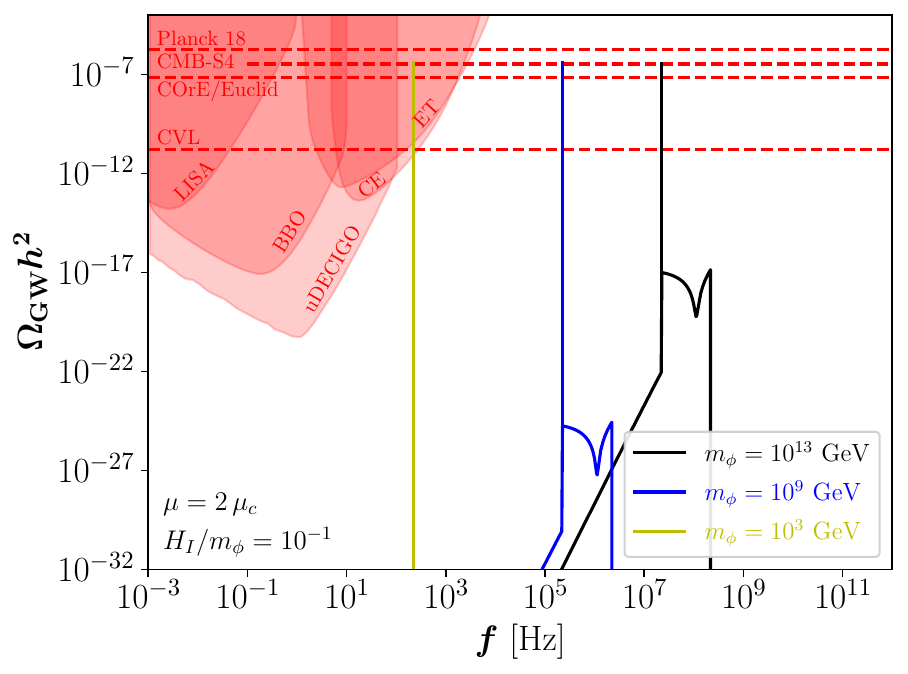}
    \includegraphics[scale=\sepf]{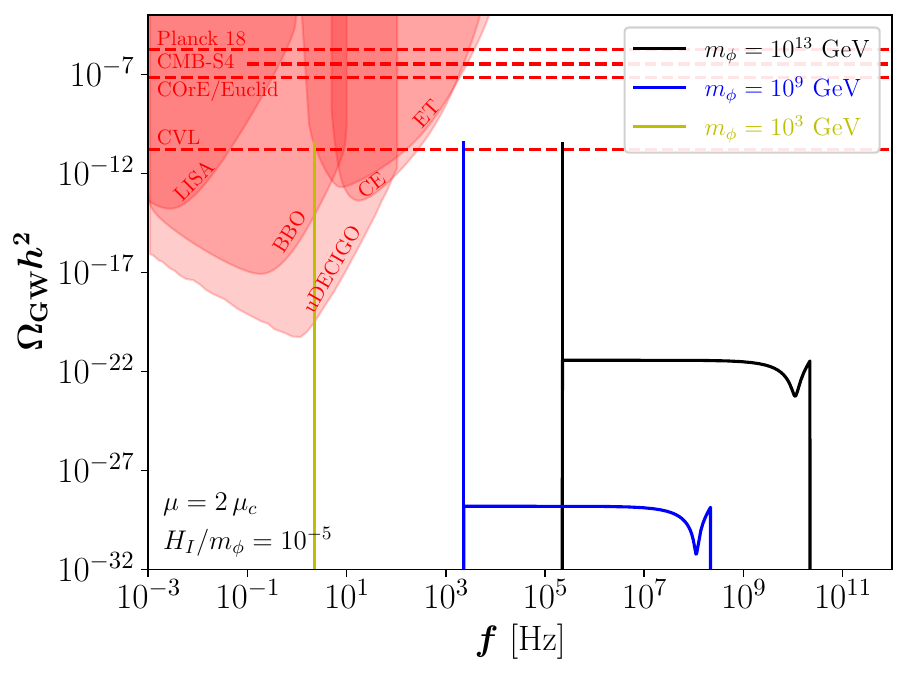}
    \caption{Top panel: Individual contributions to the GW spectrum, for $m_\phi = 10^{13}$~GeV, $\mu = 2\, \mu_c$ and $H_I/m_\phi = 10^{-3}$. Here, the subscript $B$ indicates that it is Bose-enhanced. Lower panels: Total GW spectra for $H_I/m_\phi = 10^{-1}$ (left) or $H_I/m_\phi = 10^{-5}$ (right), different inflaton masses, and $\mu = 2\, \mu_c$ (corresponding to $y/y_c = 2$). Contributions from $\phi\to\varphi\varphi h$ and $\phi\phi\to hh$ are not included due to their weak signal strength---see Eq.~\eqref{eq:GW_1-3_No_Bose} and Eq.~\eqref{eq:GW_phiphihh}.}
    \label{fig:GW-spectra}
\end{figure}
%%%%%%%%%%%%%%%%%%%%%%%%%%%%%%%%%%%%%%%%%%%%%%%
In Fig.~\ref{fig:GW-spectra}, we show the sensitivity curves of several proposed GW detectors; these include the Laser Interferometer Space Antenna (LISA)~\cite{LISA:2017pwj, Chang:2022lrw}, the Einstein Telescope (ET)~\cite{Punturo:2010zz, Hild:2010id, Sathyaprakash:2012jk, ET:2019dnz, ET:2025xjr}, the Cosmic Explorer (CE)~\cite{Reitze:2019iox}, the Big Bang Observer (BBO)~\cite{Crowder:2005nr, Corbin:2005ny, Harry:2006fi}, and the ultimate DECIGO (uDECIGO)~\cite{Seto:2001qf, Kudoh:2005as}. We note that the energy stored in the GWs behaves as dark radiation, contributing to the effective number of neutrino species, $N_\text{eff}$. In order to illustrate these constraints, we include horizontal dashed red lines in Fig.~\ref{fig:GW-spectra}. The Planck experiment provides a 95\% CL measurement of $N_{\text{eff}} = 2.99 \pm 0.34$~\cite{Planck:2018vyg}. Future experiments, such as Simons Observatory (SO)~\cite{SimonsObservatory:2018koc, SimonsObservatory:2019qwx}, CMB-S4~\cite{CMB-S4:2016ple, Abazajian:2019eic}, and CMB-HD~\cite{CMB-HD:2022bsz} are capable of probing $\Delta N_{\rm eff}$ above $0.10$, 0.06 and 0.028 at 2 $\sigma$ C.L., respectively. Some proposed experiments such as COrE~\cite{COrE:2011bfs} and Euclid~\cite{EUCLID:2011zbd} may reach $0.013$ at 2$\sigma$ C.L. In Fig.~\ref{fig:GW-spectra}, we selectively plot the limits of some of these experiments. Additionally, we include a limit of $\Delta N_{\rm eff} \lesssim 3 \times 10^{-6}$, reported in Ref.~\cite{Ben-Dayan:2019gll}, based on a hypothetical cosmic-variance-limited (CVL) CMB polarization experiment.

In the top panel of Fig.~\ref{fig:GW-spectra}, we also show examples of individual contributions to the GW spectrum for $m_\phi = 10^{13}$~GeV, $H_I/m_\phi = 10^{-3}$, and $\mu = 2\, \mu_c$ (corresponding to $y/y_c = 2$). Here, the subscript $B$ indicates that it is Bose-enhanced. In the lower panels, we show the total GW spectra for $H_I/m_\phi = 10^{-1}$ (left) or $H_I/m_\phi = 10^{-5}$ (right), different inflaton masses, and $\mu = 2\, \mu_c$. In the following, we present the calculations and discussion of each specific process.

%%%%%%%%%%%%%%%%%%%%%%%%%%%%%%%%%%%%%%%%%%%%%%%
\subsection[\texorpdfstring{$\varphi\,\varphi\to h\,h$}{φ φ → h h}]{\boldmath $\varphi\,\varphi\to h\,h$}
%%%%%%%%%%%%%%%%%%%%%%%%%%%%%%%%%%%%%%%%%%%%%%%
For gravitons produced by a pair of massless $\varphi$ particles, the squared amplitude reads~\cite{Ghiglieri:2022rfp, Bernal:2025lxp}
\begin{equation}
    \Msqr{\varphi\varphi\to hh} = \frac{2\, t^2\, (s+t)^2}{M_P^4\, s^2}\,,\label{eq:-24}
\end{equation}
where $s$ and $t$ are the Mandelstam variables. Plugging the squared amplitude into the corresponding collision term and performing the phase space integration (see Appendix~\ref{sec:integral-h}), we obtain
\begin{equation}
    \Gamma_h^{\varphi\varphi\to hh} \simeq \frac{\pi\, n_\varphi^2(a)}{30\, M_P^4\, p_h\, x_a}\, F_1(x_a)\, \Theta\left[0 < x_a < 2\right],\label{eq:-22}
\end{equation}
with $x_a \equiv (2\, p_h / m_\phi)\, a/a_I$, and
\begin{align}
    F_1(x_a) &\equiv\left(1-5x_a+10x_a^2-\frac{15x_a^3}{2}+\frac{15x_a^4}{8}-\left|1-x_a\right|^5\right) \nonumber\\
    &\simeq\frac{3}{8}\left(2-x_a\right){}^3x_a^3\thinspace,
\end{align}
where the relative error introduced in the last approximation is at the $\order{10\%}$ level, with a maximum of $20\%$ in the limits $x_a \to 0$ and $x_a \to 2$.
The derivation of Eq.~\eqref{eq:-22} has used the assumption that the production of $\varphi$ is instantaneous so that the phase space distribution of $\varphi$ can be well approximated by the Dirac delta function, i.e., $f_\varphi$ is approximately monochromatic. Note that the gravitons produced are not monochromatic, as already indicated by Eq.~\eqref{eq:-22}. 

Substituting Eq.~\eqref{eq:-22} into Eq.~\eqref{eq:f-from-int}, and for a reheaton-dominated scenario
\begin{equation}
    H(a) \simeq H_I\left(\frac{a_I}{a}\right)^2,\label{eq:-26}
\end{equation}
we obtain the present value of $f_h$:
\begin{equation}
    f_h(a_0) \simeq \frac{\pi\, a_I^2\, r_h^3\, m_\phi^2}{5\, \omega^2}\, F_1\left(\frac{2\, \omega}{a_I\, m_\phi}\right),\label{eq:-27}
\end{equation}
where 
\begin{equation}
    r_h\equiv\frac{H_I}{m_\phi}\thinspace.\label{eq:-45}
\end{equation}
Here $a_I$ can be determined by $\rho_\text{SM}(a_{\star})a_{\star}^4 = \rho_I\, a_I^4$, $\rho_\text{SM}(T) = \frac{\pi^2}{30}\, \gs(T)\, T^4$ and $\gss(T)\, T^3\, a^3 = \gss(T_0)\, T_0^3$ (entropy conservation), where $\rho_\text{SM}$ is the energy density of the SM plasma with temperature $T$,  $a_{\star}$ denotes the scale factor at the beginning of SM domination, and the subscripts ``0'' denote present values. From these relations, we obtain
\begin{equation}
    a_I \simeq \left(\frac{\pi^2}{30}\right)^{1/4} \frac{\gs(T_0)^{1/3}}{\gs^{1/12}}\frac{T_0}{\rho_I^{1/4}} \simeq 2.9 \times10^{-29} \left(\frac{H_I}{10^{13}\ \text{GeV}}\right)^{-\frac12},\label{eq:-28}
\end{equation}
where $\gs(T_0) \simeq 3.91$, $T_0 \simeq 2.73$~K, and $\gss=\gs = 106.75$. It follows that
\begin{equation}
    \ogw h^2 \simeq 1.2 \times 10^{-15} \left(\frac{f}{\text{GHz}}\right)^2 \left(\frac{m_\phi}{10^{13}~\text{GeV}}\right) r_h^2\, F_1\left(\frac{2\, \omega}{a_I\, m_\phi}\right),\label{eq:-30}
\end{equation}
where the $F_1$ part is typically of ${\cal O}(1)$. The $\Theta$ function in Eq.~\eqref{eq:-22} after cosmological redshift turns into the following interval of $f$:
\begin{equation}
    0<f<f_\text{max}\thinspace,\label{eq:-29}
\end{equation}
where 
\begin{equation}
    f_\text{max} \equiv \frac{a_I\, m_\phi}{2 \pi} \simeq 0.071~\text{GHz} \left(\frac{m_\phi}{10^{13}~\text{GeV}} \, \frac{1}{r_h}\right)^\frac12.\label{eq:-29-1}
\end{equation}
 
%%%%%%%%%%%%%%%%%%%%%%%%%%%%%%%%%%%%%%%%%%%%%%%%%%%%%%%%%%%%%%%%%%%%%%%%%%%%%%%%%%%%%%%%
\subsection[\texorpdfstring{$\phi\to\varphi\,\varphi\,h$}{ϕ → φ φ h}]{\boldmath $\phi\to\varphi\,\varphi\,h$}
%%%%%%%%%%%%%%%%%%%%%%%%%%%%%%%%%%%%%%%%%%%%%%%%%%%%%%%%%%%%%%%%%%%%%%%%%%%%%%%%%%%%%%%%
Next, consider the three-body decay $\phi \to \varphi\, \varphi\, h$, with the squared amplitude~\cite{Bernal:2025lxp} 
\begin{equation}
    \Msqr{\phi\to\varphi\varphi h} = \frac{2\, \mu^2}{M_P^2} \left(1 - \frac{m_\phi}{2\, \omega}\right)^2.\label{eq:-32}
\end{equation}
From Eq.~\eqref{eq:-32}, one can compute the corresponding GW production rate $\Gamma_h$. Its contribution of the three-body process to the total decay rate of $\phi$ is determined by
\begin{equation}
    \Gamma_\phi^{(3)} = \frac{1}{n_\phi} \int \Gamma_h(\omega)\, \frac{4 \pi\, \omega^2}{(2\pi)^3}\, d\omega\thinspace.\label{eq:-31}
\end{equation}
It is important to note that the integral in Eq.~\eqref{eq:-31} is IR divergent. This feature is very common in particle physics processes containing bremsstrahlung of soft massless bosons (e.g. $Z\to e^+e^-\gamma$, $e^-+N\to e^-+N+\gamma$). In Minkowski spacetime with zero temperature, such IR divergences are regulated by cancelation with associated loop processes---see, e.g., Ref.~\cite{Peskin:1995ev}. For gravitons, a similar cancelation has also been studied~\cite{Weinberg:1965nx}. In a finite-density environment, cancelation between emission and absorption processes can occur~\cite{Flauger:2019cam, Ai:2025xla}, if the mean free path of gravitons, which decreases with decreasing frequency in the IR limit, is sufficiently short to render the environment opaque to very soft gravitons. In the early Universe with a rapid expansion rate, QFT calculations based on Minkowski spacetime are valid only when the relevant length scale is shorter than the horizon radius $\sim1/H$. Therefore, the horizon itself provides an IR cutoff~\cite{Tokareva:2023mrt}: 
\begin{equation}
    \omega > H\thinspace.\label{eq:-33}
\end{equation}
The physical interpretation of this cutoff is clear: no waves with wavelengths longer than the size of a causally connected part of the Universe can be generated. We have checked that in our work the cutoff in Eq.~\eqref{eq:-33} is well above the scale relevant to the cancelation discussed in Refs.~\cite{Flauger:2019cam, Ai:2025xla}. Hence, Eq.~\eqref{eq:-33} will be used as the cutoff in the following calculation.
The alternative regularization via emission--absorption cancellation~\cite{Flauger:2019cam, Ai:2025xla} relies on the medium being optically thick to gravitons, such that repeated absorption and re-emission effectively regulate the infrared. This requires a graviton mean free path $\ell_{\rm mfp}(\omega)$ shorter than the Hubble scale. In the present setup, however, graviton interactions are Planck-suppressed, leading parametrically to $\ell_{\rm mfp}(\omega)\sim M_P^2/\omega^3$, which remains much larger than $H^{-1}$ throughout the relevant epoch. As a result, gravitons free-stream after production and do not undergo the multiple interactions necessary for cancellation effects to operate. The two prescriptions therefore apply to distinct physical regimes, and the horizon cutoff is the appropriate regulator in our case.

When computing the collision term of $\phi\to\varphi\,\varphi\,h$, three contributions are encountered, from double Bose-enhanced decay ($\phi\to\varphi_B\,\varphi_B\,h$,  with the subscript ``$B$'' indicating that it is Bose-enhanced), single Bose-enhanced decay ($\phi\to\varphi_B\,\varphi\,h$), and decay without any enhancement. The detailed calculations of these contributions are presented in the Appendix~\ref{sec:integral-h}. It turns out that $\phi\to\varphi_B\,\varphi_B\,h$ leads to the highest yield of gravitons produced in this work and has the strongest impact on the background among the three decay processes, as will be described in the following.

%%%%%%%%%%%%%%%%%%%%%%%%%%%%%%%%%%%%%%%%%%%%%%%%%%%%%%%%%%%%%%%%%%%%%%%%%%%%%%%%%%%%%%%%%%%%%%%%%%%%%%
\subsubsection[\texorpdfstring{$\phi\to\varphi_B\,\varphi_B\,h$}{ϕ → φB φB h}]{\boldmath $\phi\to\varphi_B\,\varphi_B\,h$}
%%%%%%%%%%%%%%%%%%%%%%%%%%%%%%%%%%%%%%%%%%%%%%%%%%%%%%%%%%%%%%%%%%%%%%%%%%%%%%%%%%%%%%%%%%%%%%%%%%%%%%
Let us start with the double Bose-enhanced contribution $\phi \to \varphi_B\, \varphi_B\, h$. Assuming that the background $\varphi$ particles is monochromatic, it is straightforward to perform the phase space integration of the collision term (see Appendix~\ref{sec:integral-h} for details) and obtain
\begin{equation}
    \Gamma_h^{\phi\to\varphi_B\varphi_Bh} \simeq \frac{4 \pi^3\, \mu^2\, n_\varphi^2\, n_\phi}{m_\phi^5\, M_P^2\, \omega^2} \left(1 - \frac{m_\phi}{2\, \omega}\right)^2 \left(\frac{a}{a_I}\right)^4 \delta\left(\omega_B - \omega\right) \Theta\left[\omega_B > H\right],\label{eq:-34}
\end{equation}
where $\omega_B \equiv m_\phi\, (1 - a_I/a)$. As indicated in Eq.~\eqref{eq:-34}, this process produces a monochromatic graviton because the energies of $\phi$ and two $\varphi_B$ are fixed. 

Substituting Eq.~\eqref{eq:-34} into Eq.~\eqref{eq:-31}, we obtain
\begin{equation}
    \Gamma_\phi^{(3)} \simeq \frac{2 \pi\, \mu^2\, n_\varphi^2}{m_\phi^5\, M_P^2} \left(\frac{a}{a_I}\right)^4 \left(1 - \frac{m_\phi}{2\, \omega_B}\right)^2 \Theta\left[\omega_B > H\right].\label{eq:-35}
\end{equation}
Note that if $a = a_I$, $\omega_B = 0$ is below $H$, leading to a vanishing result. The function $\Theta$ becomes nonzero when $a$ exceeds a certain threshold, indicated by $a_3$ and shown diagrammatically in Fig.~\ref{fig:schematic}. By solving $\omega_B=H$ with $H = H_I (a_I/a_3)^2$ and $\omega_B = m_\phi (1 - a_I/a_3)$, we get
\begin{equation}
    a_3 = a_I\, \frac{1 + \sqrt{1 + 4\, r_h}}{2}\thinspace.\label{eq:-36}
\end{equation}

When $\omega_B>H$ (i.e. $a>a_3$), one can compare $\Gamma_\phi^{(3)}$ to $H$. Assuming $r_h\ll1$, the ratio of $\Gamma_\phi^{(3)}$ to $H$ reads 
\begin{equation} \label{eq:-37}
    \frac{\Gamma_\phi^{(3)}}{H} \simeq 1.3 \times 10^2 \left(\frac{y}{y_c}\right)^2 \gg 1\,,
\end{equation}
where we have approximated $n_\varphi \simeq 2\, n_{\phi I} (a_I/a)^3$. Equation~\eqref{eq:-37} implies that once $a>a_3$ is satisfied, the remaining $\phi$ particles would decay rapidly via $\phi\to\varphi_B\,\varphi_B\,h$ (corresponding to  the ``B3'' turning point in Fig.~\ref{fig:schematic}). Since $\Gamma_\phi^{(3)}$ is at least two orders of magnitude above $H_I$, it is a good approximation to assume that this decay is instantaneous. Under this assumption,  the total energy of gravitons produced after the instantaneous decay is given by
\begin{equation}
    \rho_h^{\phi\to\varphi_B\varphi_Bh} \simeq n_{\phi I}\, Y_\phi^{({\rm stable})}\, \omega_B \left(\frac{a_I}{a}\right)^4,\label{eq:-39}
\end{equation}
where $\omega_B\simeq H_I(1-2r_h)$.
Using Eqs.~\eqref{eq:Y-phi-stable}, \eqref{eq:-21} and~\eqref{eq:-28}, we can rewrite Eq.~\eqref{eq:-39} as
\begin{equation}
    \ogw h^2 \simeq 1.8 \times 10^{-8} \left(\frac{m_\phi}{10^{13}~\text{GeV}}\, \frac{4 \times 10^{-5}}{y}\right)^2 \ln\left(\frac{32 \pi\, r_h}{y^2}\right).\label{eq:-40}
\end{equation}
This is a monochromatic spectrum (the solid vertical lines shown in Fig.~\ref{fig:GW-spectra}), with frequency $f = r_h\, f_{\max}$ where $f_{\max}$ is defined in  Eq.~\eqref{eq:-29-1}.

Equation~\eqref{eq:-39} can be reinterpreted in terms of $\Delta N_{\rm eff}$ as
\begin{equation}
    \Delta N_{\rm eff} \simeq 0.13 \left(\frac{y_c}{y}\right)^2 r_h\, \ln\left(\frac{32 \pi\, r_h}{y^2}\right).\label{eq:-41}
\end{equation}
It is interesting to note that the resulting $\Delta N_{\rm eff}$ is close to the current bound set by the Planck experiment and can be probed by future experiments. For example, if we take $m_\phi=10^{13}$~GeV, $r_h=0.1$, and $y=1.7\, y_c$, we obtain $\Delta N_{\rm eff} \simeq 0.1$, which could be reached by next-generation observatories such as CMB-S4.

%%%%%%%%%%%%%%%%%%%%%%%%%%%%%%%%%%%%%%%%%%%%%%%%%%%%%%%%%%%%%%%%%%%%%%%%%%%%%%%%%%%%%%%%%%%%%%%
\subsubsection[\texorpdfstring{$\phi\to\varphi_B\,\varphi\,h$}{ϕ → φB φ h}]{\boldmath $\phi\to\varphi_B\,\varphi\,h$}
%%%%%%%%%%%%%%%%%%%%%%%%%%%%%%%%%%%%%%%%%%%%%%%%%%%%%%%%%%%%%%%%%%%%%%%%%%%%%%%%%%%%%%%%%%%%%%%
The single Bose-enhanced decay has a lower graviton yield compared with the double Bose-enhanced decay, but it produces a continuous GW spectrum, as opposed to the monochromatic spectrum discussed above. Performing the phase space integration (see Appendix~\ref{sec:integral-h} for details) , we obtain the corresponding collision term
\begin{equation}
    \Gamma_h^{\phi\to\varphi_B\varphi h} \simeq \frac{\pi\, \mu^2\, n_\varphi\, n_\phi}{2\, m_\phi^3\, M_P^2\, \omega^2} \left(1 - \frac{m_\phi}{2\, \omega}\right)^2 \left(\frac{a}{a_I}\right)^2.\label{eq:-42}
\end{equation}
Substituting Eq.~\eqref{eq:-42} into Eq.~\eqref{eq:f-from-int}, we obtain
\begin{equation}
    f_h \simeq \frac{3 \pi\, \mu^2\, H_I\, n_{\phi I}Y_\phi^{({\rm stable})}}{8\, m_\phi^2\, \omega^4\, r_a^4} \left[\left(r_a - 1\right) \left(1 + r_a - 4\, x_a\right) + 2\, x_a^2\, \ln r_a\right],\label{eq:-43}
\end{equation}
where $r_a \equiv a/a_I$. Note that Eq.~\eqref{eq:-43} is valid only when the comoving number density of $\phi$ is constant. When $a \simeq a_3$, the remaining inflatons are rapidly depleted via $\phi\to\varphi_B\varphi_Bh$. Therefore, to compute the GW spectrum today from $\phi \to \varphi_B \varphi h$, we should cut the production at $a = a_3$ and then redshift $f_h(a = a_3)$ to the present ($a = a_0$). Using Eq.~\eqref{eq:-36} with the approximation $r_h\ll1$, we get
\begin{equation}
    f_h(a_0,\omega) \simeq \frac{3\, a_I^2\, m_\phi^2\, r_h^3}{\omega^2} \left(1 - \frac{a_I\, m_\phi}{2\, \omega}\right)^2 \ln\left(\frac{32 \pi\, r_h}{y^2}\right) \Theta\left(r_h < \frac{\omega}{a_I\, m_\phi} < \frac12\right).\label{eq:-44}
\end{equation}
The corresponding GW spectrum is
\begin{equation}
    \ogw h^2 \simeq 5.5 \times 10^{-15} \left(\frac{f}{\text{GHz}}\right)^2 \left(\frac{m_\phi}{10^{13}~\text{GeV}}\right) r_h^2 \left(\frac{f_{\max}}{2\, f} - 1\right)^2 \ln\left(\frac{32 \pi\, r_h}{y^2}\right),\label{eq:-46}
\end{equation}
and spans over the following frequency band
\begin{equation}
    r_h\, f_{\max} < f < \frac12\, f_{\max}\thinspace.\label{eq:-47}
\end{equation}

%%%%%%%%%%%%%%%%%%%%%%%%%%%%%%%%%%%%%%%%%%%%%%%%%%%%%%%%%%%%%%%%%%%%%%%%%%%%%%%%%%%%%%%%%%%%
\subsubsection[\texorpdfstring{$\phi\to\varphi\,\varphi\,h$}{ϕ → φ φ h}]{\boldmath $\phi\to\varphi\,\varphi\,h$}
%%%%%%%%%%%%%%%%%%%%%%%%%%%%%%%%%%%%%%%%%%%%%%%%%%%%%%%%%%%%%%%%%%%%%%%%%%%%%%%%%%%%%%%%%%%%
It is also possible to produce gravitons through a three-body process without any Bose enhancement. The contribution to the GW spectra is expected to be subdominant compared to the previous two cases with Bose enhancement. Indeed, the collision term for the process $\phi\to\varphi\,\varphi\,h$ is given by~\cite{Bernal:2025lxp}
\begin{equation}\label{eq:1-3_No_Bose}
    \Gamma^{\phi\to\varphi\,\varphi\,h}_h  = \frac{\mu^2\, n_\phi}{32\, \pi M_P^2\, m_\phi\, \omega} \left(1 - \frac{m_\phi}{2\, \omega}\right)^2\,,
\end{equation}
which is suppressed by a factor of $\frac{\omega m_\phi^2}{n_\varphi}$  compared to Eq.~\eqref{eq:-42} for $\phi\to\varphi_B\varphi h$. This factor is suppressed by the smallness of $\mu$, as can be seen in $\frac{\omega m_\phi^2}{n_\varphi}<\frac{m_\phi^3}{n_\varphi} < \frac{m_\phi^3}{n_{\phi I} Y_\phi^{({\rm stable})}} \sim \frac{\mu^2}{H_I m_\phi} \ll 1$.  Such suppression is also reflected in the corresponding GW spectrum. Substituting Eq.~\eqref{eq:1-3_No_Bose} into Eq.~\eqref{eq:f-from-int}, we find
\begin{align}\label{eq:fh_1-3_No_Bose}
  f_h \simeq \frac{\mu^2\, m_\phi\, n_{\phi I}Y_\phi^{({\rm stable})}}{256\, \pi\, H_I M_P^2\, \omega^3\, r_a^3} \left[\left(r_a - 1\right) \left(1 + r_a - 4\, x_a\right) + 2\, x_a^2\, \ln r_a\right]\,,
\end{align}
and the GW amplitude at present is given by
\begin{equation}\label{eq:GW_1-3_No_Bose}
    \Omega_{\text{GW}}h^2 \simeq 1.4 \times 10^{-27} \left(\frac{f}{\text{GHz}}\right)^3 \left(1- \frac{f_{\text{max}}}{2f}\right)^2 r_h^{1/2}  \left(\frac{m_\phi}{10^{13}~\text{GeV}}\right)^{5/2}  \ln\left(\frac{32 \pi\, r_h}{y^2}\right)\,.
\end{equation}
The frequency band is the same as in Eq.~\eqref{eq:-47}. It is evident that Eq.~\eqref{eq:GW_1-3_No_Bose} is significantly smaller than Eq.~\eqref{eq:-46} for the same model parameters. We also note that, for $f \ll f_{\rm max}$, the spectrum scales as $\Omega_{\rm GW} \propto f$, which is a characteristic feature of bremsstrahlung graviton production.

%%%%%%%%%%%%%%%%%%%%%%%%%%%%%%%%%%%%%%%%%%%%%%%%%%%%%%%%%%%%%%%%%%%%%%%%%%%%%%%%%%%%%%%%%%%%
\subsection[\texorpdfstring{$\varphi\phi\to\varphi h$}{φ ϕ → φ h}]{\boldmath $\varphi\phi\to\varphi h$}
%%%%%%%%%%%%%%%%%%%%%%%%%%%%%%%%%%%%%%%%%%%%%%%%%%%%%%%%%%%%%%%%%%%%%%%%%%%%%%%%%%%%%%%%%%%%
The scattering process $\varphi\phi\to\varphi h$ is expected to have a graviton yield comparable to that of the single Bose-enhanced decay $\phi\to\varphi_B\varphi h$. The Bose enhancement in the latter plays a role equivalent to the presence of an initial-state $\varphi$ with large $f_\varphi$ in the former.  Moreover, they share the same squared matrix element, which follows from crossing symmetry---see Ref.~\cite{Xu:2025wjq} for more detailed discussions. Nevertheless, a difference in the symmetry factor should be taken into account---see Tab.~1 of Ref.~\cite{Bernal:2025lxp}. The resulting collision term is
\begin{equation}
    \Gamma_h^{\varphi\phi\to\varphi h} \simeq \frac{\pi\, \mu^2\, n_\varphi\, n_\phi}{2\, m_\phi^3\, M_P^2\, \omega^2} \left(1 - \frac{m_\phi}{2\, \omega}\right)^2 \left(\frac{a}{a_I}\right)^2 \Theta\left[\frac12 < \frac{\omega}{m_\phi} < 1\right].\label{eq:-48}
\end{equation}
The corresponding $f_h$ is very similar to Eq.~\eqref{eq:-44}, except for a different $\Theta$ function:
\begin{equation}
    f_h(a=1) \simeq \frac{3\, a_I^2\, m_\phi^2\, r_h^3}{\omega^2} \left(1 - \frac{a_I\, m_\phi}{2\, \omega}\right)^2 \ln\left(\frac{32 \pi\, r_h}{y^2}\right) \Theta\left(\frac12 < \frac{\omega}{a_I\, m_\phi} < 1\right).\label{eq:-44-1}
\end{equation}
Consequently, the expression for $\ogw h^2$ in Eq.~\eqref{eq:-46} can be used with the following frequency band
\begin{equation}
    \frac12\, f_{\max} < f < f_{\max}\thinspace.\label{eq:-47-1}
\end{equation}

%%%%%%%%%%%%%%%%%%%%%%%%%%%%%%%%%%%%%%%%%%%%%%%%%%%%%%%%%%%%%%%%%%%%%%%%%%%%%%%%%%%%%%%%%%%%
\subsection[\texorpdfstring{$\phi\phi\to hh$}{ϕ ϕ → h h}]{\boldmath $\phi\phi\to hh$}
%%%%%%%%%%%%%%%%%%%%%%%%%%%%%%%%%%%%%%%%%%%%%%%%%%%%%%%%%%%%%%%%%%%%%%%%%%%%%%%%%%%%%%%%%%%%
Finally, gravitons can also be produced from inflaton pair annihilation~\cite{Ema:2020ggo, Choi:2024ilx}. The  corresponding collision term is given by~\cite{Bernal:2025lxp}
\begin{equation}\label{eq:phiphihh}
    \Gamma^{\phi\, \phi \to h\,h}_h = \frac{\pi}{16}\, \frac{n_\phi^2}{M_P^4}\, \delta(\omega - m_{\phi})\,.
\end{equation}
This channel also produces gravitons monochromatically; however, the resulting GW amplitude is typically small due to the additional Planck-scale suppression and low inflaton density drained by the Bose enhanced two-body decay. Plugging Eq.~\eqref{eq:phiphihh} into Eq.~\eqref{eq:f-from-int}, we find
\begin{equation}\label{eq:fh_phiphihh}
  f_h  \simeq \frac{ H_I\, m_\phi^{11}\, }{16 \,\pi\, M_P^4\, \mu^4\,  r_a^4\, \omega^4} \ln\left(\frac{32 \pi\, r_h}{y^2}\right)\,,
\end{equation}
for $a\lesssim a_3$. The GW amplitude at present is given by
\begin{equation}\label{eq:GW_phiphihh}
    \Omega_{\text{GW}}h^2 \simeq 2.0 \times 10^{-23} \left(\frac{m_\phi}{10^{13}~\text{GeV}}\right)^6 r_h^{-1} \left(\frac{4\times 10^{-5}}{y}\right)^4 \ln\left(\frac{32 \pi\, r_h}{y^2}\right)\,,
\end{equation}
which corresponds to a very narrow spectrum with $f\simeq f_{\text{max}}$ since the inflaton energy would be almost completely drained by the Bose enhanced three-body process at $a\simeq a_3$.  We remind the reader that in scenarios where graviton pair production from inflaton annihilation occurs during a matter-dominated phase, with $H \propto a^{-3/2}$, the resulting GW spectrum typically exhibits the scaling $\Omega_{\mathrm{GW}} h^2 \propto f^{-1/2}$, see, e.g., Refs.~\cite{Ema:2020ggo, Choi:2024ilx, Xu:2024fjl, Bernal:2025lxp, Xu:2025wjq}. In contrast, the production considered here takes place during a radiation-dominated era driven by massless $\varphi$ particles. This difference in the background expansion history explains the distinct frequency scaling of the GW spectrum compared to the results in the literature.

%%%%%%%%%%%%%%%%%%%%%%%%%%%%%%%%%%%%%%%%%%%%%%%%%%%%%%%%%%%%%%%%%%%%%%%%%%%%%%%%%%%%%%%%%%%%
\section{Conclusions \label{sec:conclusion}}
%%%%%%%%%%%%%%%%%%%%%%%%%%%%%%%%%%%%%%%%%%%%%%%%%%%%%%%%%%%%%%%%%%%%%%%%%%%%%%%%%%%%%%%%%%%%
In this work, we propose a novel reheating scenario in which the bosonic decay products of the inflaton---dubbed the reheaton---form a transient condensate with a large occupation number. Due to Bose enhancement effects, two-body inflaton decays can proceed much more efficiently than in conventional perturbative reheating scenarios, leading to a rapid transfer of energy from inflatons to radiation (reheatons), as illustrated schematically in Fig.~\ref{fig:schematic}. This enhancement accelerates the overall reheating process and modifies the inflaton decay dynamics in a way that is not captured by standard treatments. We show that once the inflaton-reheaton coupling exceeds a critical value given by Eq.~\eqref{eq:muc},  the inflaton almost immediately transfers a significant fraction of its energy to the reheaton at the beginning of reheating, as quantitatively demonstrated in Fig.~\ref{fig:critical-mu}. 

In addition to altering the standard two-body decay channels of the inflaton, we have shown that the formation of a reheaton condensate can dramatically enhance graviton production during reheating. In particular, inflaton three-body decays with a graviton in the final state are strongly amplified by Bose enhancement, owing to the large occupation number of the condensate. While such processes are ordinarily suppressed by powers of the Planck scale, we find that Bose enhancement can compensate for this suppression, rendering the corresponding decay rate comparable to---or even exceeding---the Hubble expansion rate. As a result, these three-body channels can also play a significant role in further depleting the inflaton energy density during reheating. At the same time, the enhanced graviton emission associated with these processes leads to a substantial amplification of the resulting stochastic gravitational-wave (GW) background compared to scenarios without Bose enhancement.  Moreover, the resulting GW spectra exhibit characteristic features associated with the reheating dynamics and the presence of the condensate. The corresponding stochastic GW signal can reach levels potentially accessible to future experiments, as illustrated in Fig.~\ref{fig:GW-spectra}.

In summary, we have proposed a novel perturbative reheating scenario in which quantum-statistical effects play an important role in energy transfer and graviton production. Our analysis shows that transient reheaton condensation can enhance inflaton decay rates and amplify GW emission, thereby providing an alternative mechanism that may lead to observable consequences.

%%%%%%%%%%%%%%%%%%%%%%%%%%%%%%%%%%%%%%%%%%%%%%%%%%%%%%%%%%%%%%%%%%%%%%%%%%%%%%%%%%%%%%%%%%%%
\begin{acknowledgments}
We thank Wen-Yuan Ai, Anna Tokareva, and Jing-Zhi Zhou for very useful discussions on the IR divergence and the horizon cutoff. NB thanks the Laboratoire d'Annecy-le-Vieux de Physique Théorique (LAPTh) for its hospitality and CNRS-INP for partial support. XJX thanks the high energy physics group at New York University Abu Dhabi for their hospitality---and for the opportunity to ride a camel---during a short-term visit, where part of this project was carried out. NB received funding from the grants PID2023-151418NB-I00 funded by MCIU/AEI/10.13039/501100011033/ FEDER and PID2022-139841NB-I00 of MICIU/AEI/10.13039/501100011033 and FEDER, UE. XJX is supported in part by the National Natural Science Foundation of China under grant No.~12141501 and also by the CAS Project for Young Scientists in Basic Research (YSBR-099). YX has received support from  the Natural Sciences and Engineering Research Council (NSERC) of Canada. 
\end{acknowledgments}
%%%%%%%%%%%%%%%%%%%%%%%%%%%%%%%%%%%%%%%%%%%%%%%%%%%%%%%%%%

%%%%%%%%%%%%%%%%%%%%%%%%%%%%%%%%%%%%%%%%%%%%%%%%%%%%%%%%%%%%%%%%%%%%%%%%%%%%%%%%%%%%%%%%%%%%
\appendix
%%%%%%%%%%%%%%%%%%%%%%%%%%%%%%%%%%%%%%%%%%%%%%%%%%%%%%%%%%%%%%%%%%%%%%%%%%%%%%%%%%%%%%%%%%%%
\section{Phase space integrals for the 1-to-2 process}
\label{sec:integral} 
%%%%%%%%%%%%%%%%%%%%%%%%%%%%%%%%%%%%%%%%%%%%%%%%%%%%%%%%%%%%%%%%%%%%%%%%%%%%%%%%%%%%%%%%%%%%
For $\phi\to\varphi\varphi$, which has a constant squared matrix element $|{\cal M}|^{2} = \mu^{2}$, the full expressions of the collision terms ${\cal C}[f_{\phi}]$ and ${\cal C}[f_{\varphi}]$ are
\begin{align}
    {\cal C}[f_{\phi}] & \equiv\frac{1}{2E_{\phi}}\frac{\mu^{2}}{2}\int d\Pi_{2}\,d\Pi_{3}\,(2\pi\delta)^{4}\left[-f_{\phi}\left(1+f_{2}\right)\left(1+f_{3}\right)+f_{2}f_{3}\left(1+f_{\phi}\right)\right],\label{eq:-49}\\
    {\cal C}[f_{\varphi}] & \equiv\frac{1}{2E_{\varphi}}\mu^{2}\int d\Pi_{1}\,d\Pi_{3}\,(2\pi\delta)^{4}\left[-f_{\varphi}f_{3}\left(1+f_{1}\right)+f_{1}\left(1+f_{\varphi}\right)\left(1+f_{3}\right)\right],\label{eq:-50}
\end{align}
where the subscripts $1$, $2$, and $3$ denote quantities for the 1$^\text{st}$, 2$^\text{nd}$, and 3$^\text{rd}$ particles in $\phi\to\varphi\varphi$, respectively. In the expression of ${\cal C}[f_{\varphi}]$, we assign the 2$^\text{nd}$ particle to the one considered on the left-hand side of the Boltzmann equation. The notation $(2\pi\delta)^{4}$ represents the common four-dimensional delta function responsible for momentum conservation. 

In Eq.~\eqref{eq:-49}, after expanding the products of those $f$'s and applying the symmetry of $2\leftrightarrow3$, one finds that the part in the square brackets can be replaced by $-f_{\phi}-2f_{\phi}f_{2}+f_{2}f_{3}$. In Eq.~\eqref{eq:-50}, after a similar expansion, we obtain $f_{1}+f_{1}f_{\varphi}+f_{1}f_{3}-f_{\varphi}f_{3}$. Therefore, we encounter the following phase space integrals:
\begin{align}
    I_{A} & =\frac{1}{2E_{\phi}}\frac{\mu^{2}}{2}\int d\Pi_{2}\,d\Pi_{3}\,(2\pi)^{4}\delta^{(4)}(p_{\phi}-p_{2}-p_{3})f_{\phi}\thinspace,\label{eq:-51}\\
    I_{B} & =2\times\frac{1}{2E_{\phi}}\frac{\mu^{2}}{2}\int d\Pi_{2}\,d\Pi_{3}\,(2\pi)^{4}\delta^{(4)}(p_{\phi}-p_{2}-p_{3})\,f_{\phi}f_{2}\thinspace,\label{eq:-52}\\
    I_{C} & =\frac{1}{2E_{\phi}}\frac{\mu^{2}}{2}\int d\Pi_{2}\,d\Pi_{3}\,(2\pi)^{4}\delta^{(4)}(p_{\phi}-p_{2}-p_{3})\,f_{2}f_{3}\thinspace,\label{eq:-53}\\
    I_{D} & =\frac{1}{2E_{\varphi}}\mu^{2}\int d\Pi_{1}\,d\Pi_{3}\,(2\pi)^{4}\,\delta^{(4)}(p_{1}-p_{\varphi}-p_{3})\,f_{1}\thinspace,\label{eq:-54}\\
    I_{E1} & =\frac{1}{2E_{\varphi}}\mu^{2}\int d\Pi_{1}\,d\Pi_{3}\,(2\pi)^{4}\,\delta^{(4)}(p_{1}-p_{\varphi}-p_{3})\,f_{1}f_{\varphi}\thinspace,\label{eq:-55}\\
    I_{E2} & =\frac{1}{2E_{\varphi}}\mu^{2}\int d\Pi_{1}\,d\Pi_{3}\,(2\pi)^{4}\,\delta^{(4)}(p_{1}-p_{\varphi}-p_{3})\,f_{1}f_{3}\thinspace,\label{eq:-56}\\
    I_{F} & =\frac{1}{2E_{\varphi}}\mu^{2}\int d\Pi_{1}\,d\Pi_{3}\,(2\pi)^{4}\,\delta^{(4)}(p_{1}-p_{\varphi}-p_{3})\,f_{\varphi}f_{3}\thinspace.\label{eq:-57}
\end{align}
Assuming that $f_{\phi}$ is a highly nonrelativistic distribution ($p_{\phi}\approx0$) and $\varphi$ always has its momentum $p_{\varphi}\leq m_{\phi}/2$, the results of these integrals are
\begin{align}
    I_{A} & =\frac{\mu^{2}}{32\pi\,E_{\phi}}\,f_{\phi}(p_{\phi})\,,\label{eq:-58}\\
    I_{B} & =I_{A}f_{\varphi}\left(\frac{m_{\phi}}{2}\right)\thinspace,\label{eq:-59}\\
    I_{C} & =0\thinspace,\label{eq:-60}\\
    I_{D} & =\frac{\pi\,\mu^{2}n_{\phi}}{2\,m_{\phi}^{3}}\,\delta\left(p_{\varphi}-\frac{m_{\phi}}{2}\right),\label{eq:-61}\\
    I_{E1} & =I_{E2}=\frac{\pi\,\mu^{2}n_{\phi}}{4\,m_{\phi}^{3}}\,\delta\left(p_{\varphi}-\frac{m_{\phi}}{2}\right)f_{\varphi}\left(\frac{m_{\phi}}{2}\right),\label{eq:-62}\\
    I_{F} & =0\thinspace.\label{eq:-63}
\end{align}

In the following, we present the calculations in detail. Although the specific form of $f_{\phi}$ is irrelevant to the final results as long as it is nonrelativistic, it is helpful to assume a simple analytical form of $f_{\phi}$:
\begin{equation}
    f_{\phi}=\lim_{\varepsilon\to0^{+}}2\pi^{2}\frac{n_{\phi}}{\varepsilon^{2}}\delta\left(p_{\phi}-\varepsilon\right).\label{eq:-64}
\end{equation}
Equation~\eqref{eq:-64} has been properly normalized such that 
\begin{equation}
    \int f_{\phi}(p_{\phi})\frac{d^{3}p_{\phi}}{(2\pi)^{3}}=\int\frac{n_{\phi}}{\varepsilon^{2}}\delta\left(p_{\phi}-\varepsilon\right)p_{\phi}^{2}dp_{\phi}=n_{\phi}\thinspace.\label{eq:-65}
\end{equation}

First, we compute $I_{A}$: 
\begin{align}
    I_{A} & =\frac{\mu^{2}}{4E_{\phi}}\int d\Pi_{2}\,d\Pi_{3}\,(2\pi)^{4}\,\delta^{(4)}(p_{\phi}-p_{2}-p_{3})\,f_{\phi}\nonumber \\
    & =\frac{\mu^{2}}{64\pi^{2}E_{\phi}}f_{\phi}\left.\int\frac{d^{3}\mathbf{p}_{2}}{E_{2}}\,\frac{1}{E_{3}}\,\delta(E_{\phi}-E_{2}-E_{3})\right|_{E_{3}=\sqrt{p_{\phi}^{2}+p_{2}^{2}-2p_{\phi}p_{2}c_{\theta}}}\nonumber \\
    & =\frac{\mu^{2}}{64\pi^{2}E_{\phi}}f_{\phi}\int\frac{2\pi\,p_{2}^{2}\,dp_{2}\,dc_{\theta}}{p_{2}}\,\frac{\delta\left(E_{\phi}-E_{2}-\sqrt{p_{\phi}^{2}+p_{2}^{2}-2p_{\phi}p_{2}c_{\theta}}\right)}{\sqrt{p_{\phi}^{2}+p_{2}^{2}-2p_{\phi}p_{2}c_{\theta}}}\nonumber \\
    & =\frac{\mu^{2}}{32\pi\,E_{\phi}}f_{\phi}\int\frac{p_{2}^{2}\,dp_{2}}{p_{2}}\,\frac{1}{p_{\phi}\,p_{2}}\,\Theta\left[\frac{E_{\phi}-p_{\phi}}{2}\leq p_{2}\leq\frac{E_{\phi}+p_{\phi}}{2}\right] =\frac{\mu^{2}}{32\pi\,E_{\phi}}f_{\phi}\,,\label{eq:-66}
\end{align}
where $c_{\theta}$ is the cosine of the angle between the two $\varphi$ particles. 

Next, we compute $I_{B}$, which is similar to $I_{A}$, except for an additional factor of $f_{\varphi}(p_{2})$:
\begin{align}
    I_{B} & =2\times\frac{1}{2E_{\phi}}\frac{\mu^{2}}{2}f_{\phi}\int d\Pi_{2}\,d\Pi_{3}\,(2\pi)^{4}\,\delta^{(4)}(p_{\phi}-p_{2}-p_{3})\,f_{\varphi}(p_{2})\nonumber \\
    & =2\,\frac{\mu^{2}}{32\pi\,E_{\phi}}f_{\phi}\int\frac{p_{2}^{2}\,dp_{2}}{p_{2}}\,\frac{f_{\varphi}(p_{2})}{p_{\phi}\,p_{2}}\,\Theta\left[\frac{E_{\phi}-p_{\phi}}{2}\leq p_{2}\leq\frac{E_{\phi}+p_{\phi}}{2}\right]\nonumber \\
    & =2\thinspace I_{A}\frac{1}{p_{\phi}}\int_{\frac{E_{\phi}-p_{\phi}}{2}}^{\frac{E_{\phi}+p_{\phi}}{2}}dp_{2}\,f_{\varphi}(p_{2})\thinspace.\label{eq:-67}
\end{align}
Here, the integral after $I_{A}$ can be viewed as the average value of $f_{\varphi}$ around $E_{\phi}/2$ within a small interval (the width is $p_{\phi}$). This average value approaches $f_{\varphi}\left(\frac{m_{\phi}}{2}\right)/2$ if one takes the nonrelativistic approximation in Eq.~\eqref{eq:-64} and assumes that $f_{\varphi}$ vanishes at $p_{\varphi}>m_{\phi}/2$. Then the result $I_{B}$ is given by Eq.~\eqref{eq:-59}.

Under the same assumption, the calculation of $I_{C}$ leads to a vanishing result:
\begin{align}
    I_{C} & =\frac{1}{2E_{\phi}}\frac{\mu^{2}}{2}\int d\Pi_{2}\,d\Pi_{3}\,(2\pi)^{4}\,\delta^{(4)}(p_{\phi}-p_{2}-p_{3})f_{2}f_{3}\thinspace,\nonumber \\
    & =\frac{\mu^{2}}{64\pi^{2}\,E_{\phi}}\int\frac{d^{3}\mathbf{p}_{2}}{E_{2}}\frac{d^{3}\mathbf{p}_{3}}{E_{3}}\,f_{\varphi}(p_{2})\,f_{\varphi}(p_{3})\,\delta(E_{\phi}-E_{2}-E_{3})\,\delta^{(3)}(\mathbf{p}_{\phi}-\mathbf{p}_{2}-\mathbf{p}_{3})\nonumber \\
    & =\frac{\mu^{2}}{64\pi^{2}\,E_{\phi}}\left.\int\frac{d^{3}\mathbf{p}_{2}}{E_{2}}\,\frac{f_{\varphi}(p_{2})\,f_{\varphi}(p_{3})}{E_{3}}\,\delta(E_{\phi}-E_{2}-E_{3})\right|_{\mathbf{p}_{3}=\mathbf{p}_{\phi}-\mathbf{p}_{2}}\nonumber \\
    & =\frac{\mu^{2}}{64\pi^{2}\,E_{\phi}}\int\frac{2\pi\,p_{2}^{2}\,dp_{2}\,dc_{\theta}}{p_{2}}\,\frac{f_{\varphi}(p_{2})\,f_{\varphi}(E_{\phi}-p_{2})}{p_{2}\,p_{\phi}}\,\delta\left(c_{\theta}-\frac{p_{\phi}^{2}-E_{\phi}^{2}+2E_{\phi}\,p_{2}}{2\,p_{\phi}\,p_{2}}\right) =0\thinspace.
\end{align}
The vanishing result can be physically understood from the fact that the production of $\phi$ from two $\varphi$'s requires not only $p_{\varphi}\geq m_{\phi}/2$ but also $c_{\theta}\to-1$. It can also be rigorously proved using Eq.~\eqref{eq:-64} with $\varepsilon\to0^{+}$. 

The calculation of $I_{D}$ is given as follows:
\begin{align}
    I_{D} & =\frac{1}{2E_{\varphi}}\mu^{2}\int d\Pi_{1}\,d\Pi_{3}\,(2\pi)^{4}\,\delta^{(4)}(p_{1}-p_{\varphi}-p_{3})\,f_{1}\nonumber \\
    & =\frac{\mu^{2}}{32\pi^{2}E_{\varphi}}\left.\int\frac{d^{3}\mathbf{p}_{1}}{E_{1}}\,\frac{1}{p_{3}}\,\delta(E_{1}-p_{\varphi}-p_{3})\,f_{\phi}(p_{1})\right|_{\mathbf{p}_{3}=\mathbf{p}_{1}-\mathbf{p}_{\varphi}}\nonumber \\
    & =\frac{\mu^{2}}{32\pi^{2}E_{\varphi}}\int\frac{2\pi p_{1}^{2}dp_{1}\,dc_{\theta}}{E_{1}}\,\frac{\delta\left(E_{1}-p_{\varphi}-\sqrt{p_{1}^{2}+p_{\varphi}^{2}-2p_{1}p_{\varphi}c_{\theta}}\right)}{\sqrt{p_{1}^{2}+p_{\varphi}^{2}-2p_{1}p_{\varphi}c_{\theta}}}f_{\phi}(p_{1})\nonumber \\
    & =\frac{\mu^{2}}{16\pi\,E_{\varphi}}\int\frac{p_{1}^{2}dp_{1}}{E_{1}}\,\frac{1}{p_{1}p_{\varphi}}\,f_{\phi}(p_{1})\,\Theta\left[\frac{m_{\phi}^{2}-4p_{\varphi}^{2}}{4p_{\varphi}}\leq p_{1}\right]\nonumber \\
    & =\frac{\mu^{2}}{16\pi\,E_{\varphi}^{2}}\int_{0}^{\infty}\frac{p_{1}dp_{1}}{E_{1}}\,\frac{2\pi^{2}n_{\phi}}{\varepsilon^{2}}\delta(p_{1}-\varepsilon)\,\Theta\left[\frac{m_{\phi}^{2}-4p_{\varphi}^{2}}{4p_{\varphi}}\leq p_{1}\right]\nonumber \\
    & =\frac{\mu^{2}}{16\pi\,E_{\varphi}^{2}}\frac{2\pi^{2}n_{\phi}}{\sqrt{m_{\phi}^{2}+\varepsilon^{2}}}\,\frac{1}{\varepsilon}\Theta\left[\left|p_{\varphi}-\frac{1}{2}\sqrt{m_{\phi}^{2}+\varepsilon^{2}}\right|\le\frac{\varepsilon}{2}\right]\nonumber \\
    & =\frac{\pi\,\mu^{2}\,n_{\phi}}{2\,m_{\phi}^{3}}\,\delta\left(p_{\varphi}-\frac{m_{\phi}}{2}\right),\label{eq:-68}
\end{align}
where $c_{\theta}$ is the cosine of the angle between $\vec{p}_{1}$ and $\vec{p}_{\varphi}$. 

The calculation of $I_{E1}$ is simple since $f_{\varphi}$ can be factored out of the integral. So, the result is expected to be similar to Eq.~\eqref{eq:-68}. However, one needs to be careful about a factor of $1/2$ arising from that the delta function in Eq.~\eqref{eq:-68} is symmetric with respect to $p_{\varphi}=m_{\phi}/2$, while $f_{\varphi}$ vanishes if $p_{\varphi}>m_{\phi}/2$. 

If $f_{\varphi}$ is replaced by $f_{3}=f_{\varphi}(p_{3})$, $I_{E1}$ changes to $I_{E2}$. The calculation is different, but it leads to the same result: 
\begin{align}
    I_{E2} & =\frac{\mu^{2}}{2E_{\varphi}}\int d\Pi_{1}\,d\Pi_{3}\,(2\pi)^{4}\,\delta^{(4)}(p_{1}-p_{\varphi}-p_{3})\,f_{1}\,f_{3}\nonumber \\
    & =\frac{\mu^{2}}{32\pi^{2}E_{\varphi}}\left.\int\frac{d^{3}\mathbf{p}_{1}}{E_{1}}\,\frac{1}{p_{3}}\,\delta(E_{1}-p_{\varphi}-p_{3})\,f_{\phi}(p_{1})\,f_{\varphi}(p_{3})\right|_{\mathbf{p}_{3}=\mathbf{p}_{1}-\mathbf{p}_{\varphi}}\nonumber \\
    & =\frac{\mu^{2}}{16\pi\,E_{\varphi}}\int\frac{p_{1}^{2}dp_{1}}{E_{1}}\,\frac{1}{p_{1}p_{\varphi}}\,f_{\phi}(p_{1})\,f_{\varphi}(E_{1}-p_{\varphi})\,\Theta\left[\frac{m_{\phi}^{2}-4p_{\varphi}^{2}}{4p_{\varphi}}\leq p_{1}\right]\nonumber \\
    & =\frac{\mu^{2}}{16\pi\,E_{\varphi}^{2}}\int_{0}^{\infty}\frac{p_{1}\,dp_{1}}{E_{1}}\,f_{\phi}(p_{1})\,f_{\varphi}(E_{1}-p_{\varphi})\,\Theta\left[\frac{m_{\phi}^{2}-4p_{\varphi}^{2}}{4p_{\varphi}}\leq p_{1}\right]\nonumber \\
    & =\frac{\mu^{2}}{16\pi\,E_{\varphi}^{2}}\int_{0}^{\infty}\frac{p_{1}\,dp_{1}}{E_{1}}\,\frac{2\pi^{2}n_{\phi}}{\varepsilon^{2}}\delta(p_{1}-\varepsilon)\,f_{\varphi}(E_{1}-p_{\varphi})\,\Theta\left[\frac{m_{\phi}^{2}-4p_{\varphi}^{2}}{4p_{\varphi}}\leq p_{1}\right]\nonumber \\
    & =\frac{\mu^{2}}{16\pi\,E_{\varphi}^{2}}\frac{2\pi^{2}n_{\phi}}{\sqrt{m_{\phi}^{2}+\varepsilon^{2}}}\,f_{\varphi}\left(\sqrt{m_{\phi}^{2}+\varepsilon^{2}}-p_{\varphi}\right)\,\frac{1}{\varepsilon}\Theta\left[\left|p_{\varphi}-\frac{1}{2}\sqrt{m_{\phi}^{2}+\varepsilon^{2}}\right|\le\frac{\varepsilon}{2}\right]\nonumber \\
    & =\frac{\pi\,\mu^{2}n_{\phi}}{4\,m_{\phi}^{3}}\,\delta\left(p_{\varphi}-\frac{m_{\phi}}{2}\right)f_{\varphi}\left(\frac{m_{\phi}}{2}\right)\,,
\end{align}
where we need to take into account a similar factor of $1/2$ as before. 

The calculation of $I_{F}$ is similar to that of $I_{C}$, which leads to a vanishing result.

%%%%%%%%%%%%%%%%%%%%%%%%%%%%%%%%%%%%%%%%%%%%%%%%%%%%%%%%%%%%%%%%%%%%%%%%%%%%%%%%%%%%%5
\section{\boldmath Solving the Boltzmann equation of \texorpdfstring{$f_\varphi$}{fφ}} \label{sec:solve-f-R}
%%%%%%%%%%%%%%%%%%%%%%%%%%%%%%%%%%%%%%%%%%%%%%%%%%%%%%%%%%%%%%%%%%%%%%%%%%%%%%%%%%%%%5
The unintegrated Boltzmann equation of $f_{\varphi}$ reads
\begin{equation}
    \left(\frac{\partial}{\partial t}-Hp_{\varphi}\frac{\partial}{\partial p_{\varphi}}\right)f_{\varphi}(t,p_{\varphi}) = \frac{16\pi^{2}\Gamma_{\phi}^{(0)}n_{\phi}}{m_{\phi}^{2}}\, \delta\left(p_{\varphi}-\frac{m_{\phi}}{2}\right) {\cal B}_{e}\thinspace.\label{eq:Boltz-f-r}
\end{equation}
Since ${\cal B}_{e}$ is multiplied by the $\delta$ function, one can replace ${\cal B}_{e} = 1 + f_{\varphi} \left(t, \frac{m_{\phi}}{2}\right) \to 1 + f_{\varphi} \left(t, p_{\varphi}\right)$ and rewrite Eq.~\eqref{eq:Boltz-f-r} as 
\begin{equation}
    \left(\frac{\partial}{\partial t}-Hp_{\varphi}\frac{\partial}{\partial p_{\varphi}}\right)F(t,p_{\varphi})=\frac{16\pi^{2}\Gamma_{\phi}^{(0)}n_{\phi}}{m_{\phi}^{2}}\delta\left(p_{\varphi}-\frac{m_{\phi}}{2}\right),\label{eq:-16}
\end{equation}
with 
\begin{equation}
    F(t,p_{\varphi})\equiv\ln\left[1+f_{\varphi}(t,p_{\varphi})\right].\label{eq:-17}
\end{equation}
Next, it is straightforward to apply Eq.~(B.3) in Ref.~\cite{Wu:2024uxa}. This allows us to obtain $F$ analytically:
\begin{align}
    F & =\int_{a_{I}}^{a}da'\,\frac{16\pi^{2}\Gamma_{\phi}^{(0)}n_{\phi}(a')}{m_{\phi}^{2}\,a'\,H(a')}\,\delta\left(p_{\varphi}\,\frac{a}{a'}-\frac{m_{\phi}}{2}\right)\nonumber \\
    & =\int_{a_{I}}^{a}da'\,\frac{16\pi^{2}\Gamma_{\phi}^{(0)}n_{\phi}(a')}{m_{\phi}^{2}\,a'\,H(a')}\,\frac{{a'}^{2}}{a\,p_{\varphi}}\,\delta\left(a'-\frac{2\,p_{\varphi}}{m_{\phi}}\,a\right)\nonumber \\
    & =\frac{32\pi^{2}\Gamma_{\phi}^{(0)}n_{\phi}\left(\frac{2\,p_{\varphi}}{m_{\phi}}\,a\right)}{m_{\phi}^{3}\,H\left(\frac{2\,p_{\varphi}}{m_{\phi}}\,a\right)}\,\Theta\left[\frac{m_{\phi}}{2}\,\frac{a_{I}}{a}\leq p_{\varphi}\leq\frac{m_{\phi}}{2}\right].\label{eq:-69}
\end{align}
From Eq.~\eqref{eq:-69} and Eq.~\eqref{eq:-17}, we obtain the solution in Eq.~\eqref{eq:f-reheaton}.

%%%%%%%%%%%%%%%%%%%%%%%%%%%%%%%%%%%%%%%%%%%%%%%%%%%%%%%%%%%%%%%%%%%%%%%%%%%%%%%%%%%%%%%%%%%%%
\section{Phase space integrals for graviton production processes \label{sec:integral-h}}
%%%%%%%%%%%%%%%%%%%%%%%%%%%%%%%%%%%%%%%%%%%%%%%%%%%%%%%%%%%%%%%%%%%%%%%%%%%%%%%%%%%%%%%%%%%%%
For $\varphi\varphi\to hh$, the collision term reads 
\begin{equation}
    \Gamma_{h}=\frac{1}{2\omega}\int d\Pi_{1}d\Pi_{2}d\Pi_{3}\,f_{1}f_{2}\,\frac{1}{2}\thinspace\frac{2t^{2}(s+t)^{2}}{M_{P}^{4}\,s^{2}}\,(2\pi)^{4}\,\delta^{(4)}(p_{1}+p_{2}-p_{3}-p_{4})\thinspace,\label{eq:-70}
\end{equation}
where we have included a symmetry factor of $1/2$. 

By inserting the following two identities into Eq.~\eqref{eq:-70}:
\begin{align}
    1 & =\int\frac{d^{4}q}{(2\pi)^{4}}(2\pi)^{4}\delta^{(4)}\left(q-p_{1}-p_{2}\right),\label{eq:-73}\\
    1 & =\int\frac{ds}{2\pi}(2\pi)\delta\left(s+q^{2}\right),\label{eq:-74}
\end{align}
we obtain
\begin{align}
    \Gamma_{h} & =\frac{1}{4\omega}\int\frac{ds}{2\pi}\int\frac{d^{4}q}{(2\pi)^{4}}\int d\Pi_{1}\,d\Pi_{2}\,f_{1}f_{2}\,(2\pi)^{4}\delta^{(4)}(p_{1}+p_{2}-q)\nonumber \\
    & \qquad \times\int d\Pi_{3}(2\pi)^{4}\delta^{(4)}(q-p_{3}-p_{4})\frac{2t^{2}(s+t)^{2}}{M_{P}^{4}\,s^{2}}\nonumber \\
    & =\frac{1}{16(2\pi)^{2}\omega^{2}}\int_{\omega}^{\infty}dE_{q}\int_{0}^{s_{\max1}}ds\int d\Pi_{1}\,d\Pi_{2}\,f_{1}f_{2}\,(2\pi)^{4}\delta^{(4)}(p_{1}+p_{2}-q)\frac{2t^{2}(s+t)^{2}}{M_{P}^{4}\,s^{2}}\nonumber \\
    & =\frac{1}{64(2\pi)^{4}\omega^{2}}\int_{\omega}^{\infty}dE_{q}\int_{0}^{E_{q}}dp_{1}f_{\varphi}(p_{1})f_{\varphi}(E_{q}-p_{1})\nonumber \\
    &\qquad \times\int_{0}^{s_{\max2}}\frac{ds}{\sqrt{E_{q}^{2}-s}}\int_{0}^{2\pi}d\varphi\frac{2t^{2}(s+t)^{2}}{M_{P}^{4}\,s^{2}}\thinspace,\label{eq:-75}
\end{align}
where $s_{\max1}\equiv4\omega(E_{q}-\omega)$, $s_{\max2} \equiv \min \left[s_{\max1}, 4E_{1}(E_{q}-E_{1})\right]$, and $\varphi$ is the azimuthal angle between the planes of $(\mathbf{q},\mathbf{p}_{1})$ and $(\mathbf{q},\mathbf{p}_{4})$.

Next, we use the monochromatic phase-space distribution of $\varphi$:
\begin{equation}
    f_{\varphi}(p)=2\pi^{2}\frac{n_{\varphi}}{p_{\varphi}^{2}}\delta\left(p-p_{\varphi}\right),\label{eq:-64-1}
\end{equation}
where $p_{\varphi}=a_{I}m_{\phi}/(2a)$ and $n_{\varphi}$ scales as $a^{-3}$. With the delta function in Eq.~\eqref{eq:-64-1}, it is straightforward to proceed with the integration and obtain Eq.~\eqref{eq:-22}.

For $\phi\to\varphi_{B}\varphi_{B}h$, the collision term reads
\begin{equation}
    \Gamma_{h}=\frac{1}{2\omega}\frac{2\mu^{2}}{M_{P}^{2}}\left(1-\frac{m_{\phi}}{2\omega}\right)^{2}\int d\Pi_{1}d\Pi_{2}d\Pi_{3}\,f_{1}f_{2}f_{3}\,(2\pi)^{4}\,\delta^{(4)}(p_{1}-p_{2}-p_{3}-p_{4})\thinspace,\label{eq:-70-1}
\end{equation}
where we have factored the squared matrix element out of the integral. 

Since $f_{1}$, $f_{2}$, and $f_{3}$ in the integral are all monochromatic [see Eqs.~\eqref{eq:-64} and~\eqref{eq:-64-1}], we integrate out the corresponding delta functions and get
\begin{equation}
    \Gamma_{h}=\frac{1}{2\omega}\frac{2\mu^{2}}{M_{P}^{2}}\left(1-\frac{m_{\phi}}{2\omega}\right)^{2}\frac{n_{\phi}n_{\varphi}^{2}}{8m_{\phi}p_{\varphi}^{2}}\int\frac{d\Omega_{2}}{4\pi}\frac{d\Omega_{3}}{4\pi}\,(2\pi)^{4}\,\delta(m_{\phi}-2p_{\varphi}-\omega)\delta^{(3)}(p_{\varphi}\mathbf{n}_{2}+p_{\varphi}\mathbf{n}_{3}+\omega\mathbf{n}_{4})\thinspace,\label{eq:-72}
\end{equation}
where $\vec{n}_{2,3,4}$ denote unit vectors of the 2$^\text{nd}$, 3$^\text{rd}$ and
4$^\text{th}$ particles of the process. 

Without loss of generality, we assume that $\vec{n}_{4}$ aligns with the $z$ axis and parameterize $\vec{n}_{2,3}$ as follows.
\begin{equation}
    \vec{n}_{i}=(s_{i}\cos\phi_{i},\ s_{i}\sin\phi_{i},\ c_{i})\ \ {\rm with}\ \ i\in\{2,3\}\thinspace,\label{eq:-71}
\end{equation}
where $(s_{i},\ c_{i})\equiv(\sin\theta_{i},\ \cos\theta_{i})$. 

With the above parameterization, the angular part of the phase space integral can be performed explicitly:
\begin{align}
    &\int\frac{d\Omega_{2}}{4\pi}\frac{d\Omega_{3}}{4\pi}\delta^{(3)}(p_{\varphi}\vec{n}_{2}+p_{\varphi}\vec{n}_{3}+\omega\vec{n}_{4})\nonumber \\
    &\quad =  \int\frac{dc_{2}d\phi_{2}}{4\pi}\frac{dc_{3}d\phi_{2}}{4\pi}\frac{1}{p_{\varphi}^{3}}\delta\left(\sum_{i}s_{i}\cos\phi_{i}\right)\delta\left(\sum_{i}s_{i}\sin\phi_{i}\right)\delta\left(\frac{\omega}{p_{\varphi}}+\sum_{i}c_{i}\right) = \frac{1}{8\pi p_{\varphi}^{2}\omega}\thinspace.\label{eq:-76}
\end{align}
Therefore, Eq.~\eqref{eq:-72} proceeds as follows
\begin{align}
    \Gamma_{h} & =\frac{1}{2\omega}\frac{2\mu^{2}}{M_{P}^{2}}\left(1-\frac{m_{\phi}}{2\omega}\right)^{2}\frac{n_{\phi}n_{\varphi}^{2}}{8m_{\phi}p_{\varphi}^{2}}(2\pi)^{4}\frac{1}{8\pi p_{\varphi}^{2}\omega}\,\delta(m_{\phi}-2p_{\varphi}-\omega)\nonumber \\
    & =\frac{4\pi^{3}\mu^{2}n_{\varphi}^{2}n_{\phi}}{m_{\phi}^{5}M_{P}^{2}\omega^{2}}\left(1-\frac{m_{\phi}}{2\omega}\right)^{2}\left(\frac{a}{a_{I}}\right)^{4}\delta\left(\omega_{B}-\omega\right).\label{eq:-78}
\end{align}
For $\phi\to\varphi_{B}\varphi h$, the collision term reads
\begin{equation}
    \Gamma_{h}=\frac{1}{2\omega}\frac{2\mu^{2}}{M_{P}^{2}}\left(1-\frac{m_{\phi}}{2\omega}\right)^{2}\int d\Pi_{1}d\Pi_{2}d\Pi_{3}\,f_{1}f_{2}\,(2\pi)^{4}\,\delta^{(4)}(p_{1}-p_{2}-p_{3}-p_{4})\thinspace,\label{eq:-77}
\end{equation}
similar to Eq.~\eqref{eq:-70-1}, but without $f_{3}$. Integrating the delta functions in $f_{1}$ and $f_{2}$, we get
\begin{equation}
    \Gamma_{h}=\frac{1}{2\omega}\frac{2\mu^{2}}{M_{P}^{2}}\left(1-\frac{m_{\phi}}{2\omega}\right)^{2}\frac{n_{\phi}n_{\varphi}}{4m_{\phi}p_{\varphi}}\int\frac{d\Omega_{2}}{4\pi}d\Pi_{3}\,(2\pi)^{4}\delta^{(4)}(p_{1}-p_{2}-p_{3}-p_{4})\,,\label{eq:-72-1}
\end{equation}
where $\vec{n}_{2,3,4}$ denote unit vectors of the 2$^\text{nd}$, 3$^\text{rd}$ and
4$^\text{th}$ particles of the process. Since particle $3$ is massless, we
have
\begin{equation}
    \int d\Pi_{3}\,(2\pi)^{4}\delta^{(4)}(p_{1}-p_{2}-p_{3}-p_{4})=2\pi\frac{1}{2E_{124}}\delta(E_{124}-p_{124})\thinspace,\label{eq:-79}
\end{equation}
where $E_{124}\equiv|E_{1}-E_{2}-E_{4}|$ and $p_{124}\equiv|\vec{p}_{1}-\vec{p}_{2}-\vec{p}_{4}|$. Next, we perform the angular part of the phase space integral
\begin{align}
    &\int\frac{d\Omega_{2}}{4\pi}d\Pi_{3}\,(2\pi)^{4}\delta^{(4)}(p_{1}-p_{2}-p_{3}-p_{4})\nonumber \\
    &\qquad =\int\frac{d\Omega_{2}}{4\pi}2\pi\frac{1}{2E_{124}}\delta(E_{124}-p_{124}) =\frac{\pi}{2E_{2}\omega}\thinspace.\label{eq:-80}
\end{align}
Substituting Eq.~\eqref{eq:-80} into Eq.~\eqref{eq:-72-1}, we obtain
\begin{align}
    \Gamma_{h} & =\frac{1}{2\omega}\frac{2\mu^{2}}{M_{P}^{2}}\left(1-\frac{m_{\phi}}{2\omega}\right)^{2}\frac{n_{\phi}n_{\varphi}}{4m_{\phi}p_{\varphi}}\frac{\pi}{2p_{\varphi}\omega}\nonumber \\
    & =\frac{\pi n_{\varphi}n_{\phi}}{2m_{\phi}^{3}\omega^{2}}\frac{\mu^{2}}{M_{P}^{2}}\left(1-\frac{m_{\phi}}{2\omega}\right)^{2}\left(\frac{a}{a_{I}}\right)^{2}.\label{eq:-81}
\end{align}

%%%%%%%%%%%%%%%%%%%%%%%%%
\bibliographystyle{JHEP}
\bibliography{biblio}
%%%%%%%%%%%%%%%%%%%%%%%%%
\end{document}